\documentclass[journal]{IEEEtran}
\IEEEoverridecommandlockouts
\usepackage{amsfonts}
\usepackage{amsmath}
\usepackage{amssymb}
\usepackage{caption}
\usepackage{authblk}
\usepackage{bbm}
\usepackage{graphicx}
\usepackage{epsfig}
\usepackage{subfigure}
\usepackage{stfloats}
\usepackage{eqnarray}
\usepackage{makecell}
\usepackage{multirow}
\usepackage{enumerate}
\usepackage{bookmark}
\usepackage{cite}
\usepackage{balance}
\usepackage{color}
\usepackage{bm}
\usepackage{algorithm}
\usepackage{algpseudocode}
\usepackage{url}
\usepackage{multirow}
\usepackage{longtable}
\usepackage{romannum}
\usepackage{diagbox}
\usepackage{tablefootnote}
\usepackage[justification=centering]{caption}
\newtheorem{theorem}{\textbf{Theorem}}

\usepackage{amsmath,mathtools,amssymb}
\DeclareMathOperator{\sign}{sign}
\DeclarePairedDelimiter{\norm}{\lVert}{\rVert}

\newcommand{\normmm}[1]{{\left\vert\kern-0.25ex\left\vert\kern-0.25ex\left\vert #1 
   \right\vert\kern-0.25ex\right\vert\kern-0.25ex\right\vert}}

\newcommand{\bb}[1]{#1}

\begin{document}
\pagenumbering{arabic}

\title{Nonconvex Regularized Gradient Projection Sparse Reconstruction for Massive MIMO Channel Estimation} 

\author{Pengxia~Wu and
             Julian~Cheng,~\IEEEmembership{Senior~Member,~IEEE}
\thanks{P. Wu and J. Cheng are with the School of Engineering, The University
of British Columbia, Kelowna, BC V1X 1V7, Canada (e-mail:pengxia.wu@alumni.ubc.ca, julian.cheng@ubc.ca). This work was supported by an NSERC Discovery Grant.}
}

\maketitle

\begin{abstract}
	Novel sparse reconstruction algorithms are proposed for beamspace channel estimation in massive multiple-input multiple-output systems. The proposed algorithms minimize a least-squares objective having a nonconvex regularizer. This regularizer removes the penalties on a few \mbox{large-magnitude} elements from the conventional \mbox{$\ell_1$-norm} regularizer, and thus it only forces penalties on the remaining elements that are expected to be zeros. Accurate and fast reconstructions can be achieved by performing gradient projection updates within the framework of difference of convex functions (DC) programming. A double-loop algorithm and a single-loop algorithm are proposed via different DC decompositions, and these two algorithms have distinct computational complexities and convergence rates. An extension algorithm is further proposed by designing new step sizes for the single-loop algorithm. The extension algorithm has a faster convergence rate and can achieve approximately the same level of accuracy as the proposed double-loop algorithm. Numerical results show signiﬁcant advantages of the proposed algorithms over existing reconstruction algorithms in terms of reconstruction accuracies and runtimes. Compared with the benchmark channel estimation approaches, the proposed algorithms can achieve smaller channel reconstruction error and higher achievable spectral efficiency.   
%

\end{abstract}

\IEEEpeerreviewmaketitle
\section{Introduction}
	As a key technology for the ﬁfth-generation and beyond communication systems, massive multiple-input multiple-output (MIMO) heavily relies on accurate knowledge of channel state information (CSI) to reap potential performance beneﬁts from a large number of antennas.
	However, acquiring downlink CSI in a resource-efficient manner is problematic for frequency division duplex (FDD) systems where uplink and downlink channels operate in different spectrum bands. 
	For such a system, downlink channels are estimated at the user equipments (UEs) and then the UEs send back the estimated CSI to the base station (BS); both the overheads of downlink pilot and CSI feedback are proportional to the number of BS antennas, which can be large for massive MIMO systems.
	To reduce the overhead of downlink CSI acquisition, one can develop sparse channel estimation methods by exploiting channel sparsity in certain domains.
	More specifically, the unknown massive MIMO channels can be represented as high-dimensional sparse vectors in certain basis \cite{bajwa2010compressed}.
	Based on valid sparse representations of massive MIMO channels, transmitting pilots over the channel can be regarded as a linear mapping of sparse channel vectors onto a compact subspace to obtain lower-dimensional observations.
	Then, the full-dimensional channel vectors are expected to be reconstructed from the lower-dimensional observations.
	In such sparse channel estimation methods, the length of pilot sequence for each antenna is no longer proportional to the number of BS antennas, but depends on the sparsity level of the channel. Thus, the length of downlink pilot sequences can be much less than the number of BS antennas, and the CSI acquisition overhead for FDD massive MIMO will be substantially reduced. 

	Sparse channel estimation \cite{sparse2002Cotter} has been proposed for many traditional communication applications such as orthogonal frequency-division multiplexing systems \cite{improving2005Raghavendra}, ultra-wideband communications \cite{ultra2007Jose}, pulse-shaping multicarrier systems \cite{compressive2020Georg} and underwater acoustic communications  \cite{sparse2010Berger}. 
	 The channels in these applications typically have sparse impulse responses because the corresponding multipath channel has a large delay spread and a small number of nonzero taps \cite{sparse2002Cotter}.
	Nowadays, sparse channel estimation has been increasingly investigated for massive MIMO systems.
	Due to a limited number of scattering clusters and the small angular spread for each scattering cluster, the massive MIMO channel in \mbox{angular-domain} (beamspace) exhibits a sparse feature \cite{brady2013beamspace}, i.e., the majority of channel energy is occupied by a few dimensions and most channel coefficients are either zero or nearly zero.
	Various compressive sensing-based beamspace channel estimation schemes have been developed for massive MIMO \cite{gao2015spatially, gao2016structured, Berger2010application,eltayeb2014compressive,choi2015downlink,tseng2016enhanced,almosa2018downlink,shen2016joint,rao2014distributed,liu2017closed} and millimeter-wave (mmWave) massive MIMO \cite{gao2017reliable,gao2016channel_letter,Venugopal2017,Lee2014Globecom, rod2018frequency,coma2018channel,Mo2018,Dai2016Beamspace,Yang2018TCOM}.
	Several orthogonal matching pursuit (OMP) based algorithms have been proposed for sparse channel estimations of mmWave massive MIMO \cite{Venugopal2017,Lee2014Globecom, rod2018frequency, coma2018channel}.
	The quantized compressive sensing technique has been applied to the channel estimation for mmWave massive MIMO with few-bit analog-to-digital converters \cite{Mo2018}.
	The image reconstruction technique has been applied to mmWave MIMO channel estimation to develop an SCAMPI \footnote{The SCAMPI algorithm is the sparse noninformative parameter estimator-based cosparse analysis approximate message passing for imaging algorithm \cite{Yang2018TCOM}.} algorithm based on the approximate message passing algorithm \cite{Yang2018TCOM}.
	
	Apart from the compressive sensing-based channel estimation scheme, there are other channel estimation schemes such as the codebook-based method \cite{Alkehateeb2014}, the parametric channel estimation \cite{Wang2019JSTSP} and the subspace decomposition-based channel estimation \cite{Zhang2020TVT, Buzzi2017}.
	The codebook-based beamforming \cite{Alkehateeb2014} only estimates the CSI implicitly; however, explicit CSI is often required for general beamforming tasks.
	The parametric channel estimation can achieve super-resolutional channel estimation without a discrete angular-domain channel assumption, but it requires the array manifold as a priori.
	Moreover, real-time CSI acquisition is challenging when using the parametric channel estimation in a high-mobility environment, since it is time-consuming to acquire the angle of arrivals (AoAs), angle of departures (AoDs) and path gains of all channel paths.
	The subspace decomposition-based channel estimation requires iterative channel sounding operations between the BS and UEs, and this echo process accumulates noise and thus can degrade channel estimation accuracy.	
	
	Among various channel estimation methods, the compressive sensing-based channel estimation is attractive since it can obtain instantaneous CSI based on the channel sparsity feature, without requiring statistical priori. 
	Therefore, we focus on the compressive sensing-based scheme and aim to develop efficient algorithms for sparse channel reconstruction in massive MIMO systems.
	Sparse channel reconstruction requires the reconstruction of high-dimensional sparse channel vectors from lower-dimensional pilot observations.
	The ultimate sparse reconstruction problem is an NP-hard \mbox{$\ell_0$-minimization} optimization problem \cite{Thi2018DC}.
	One approach is to replace the \mbox{$\ell_0$-norm} by other simple functions to seek approximate solutions.
	For this approach, the most popular method is to use the $\ell_1$-norm to approximate the $\ell_0$-norm, and this method is known as $\ell_1$-relaxation or convex relaxation.
	Various well-known $\ell_1$-relaxation algorithms have been proposed such as the $l1\_ls$ \cite{Kim2007l1ls}, Bregman iterative regularization \cite{Yin2008Bregman}, gradient projection for sparse reconstruction (GPSR) \cite{figueiredo2007gradient}, sparse reconstruction by separable approximation (SpaRSA) \cite{wright2009sparse}, iterative shrinkage-thresholding algorithm (ISTA) \cite{blumensath2008thomas} and the fast iterative shrinkage-thresholding algorithm (FISTA) \cite{Beck2009Afast}.
	These $\ell_1$-relaxation algorithms solve convex optimizations to guarantee global optimal solutions; however, an $\ell_1$-norm is a loose approximation of $\ell_0$-norm, leading to biased estimates.
	To achieve tighter approximations, nonconvex functions have been proposed such as the $\ell_q$-norm for $0 \le q \le 1$ \cite{Foucart2009} and the difference of two continuous functions \cite{Gasso2009}. 
	Another popular method is the greedy approach, including some representative algorithms such as OMP \cite{pati1993orthogonal, tropp2007signal}, CoSaMP \cite{needell2009CoSaMP} and the least angle regression (LARS) \cite{efron2004least}.
	These greedy methods work well for sufficiently sparse vectors, but the reconstruction accuracies and speeds degrade severely when the sparsity of reconstructing vectors reduces.
	
	In this paper, we propose to use the classic difference of convex functions (DC) programming and gradient projection descent algorithm to solve the fundamental sparse recovery problem.
	We develop novel algorithms that are applied to beamspace channel estimation for massive MIMO systems.
	Instead of approximating the \mbox{$\ell_0$-norm}, we represent exactly the \mbox{$\ell_0$-norm} constraint by introducing the \mbox{top-$(K,1)$} norm \cite{Gotoh2018DC}\footnote{\bb{The top-($K$,1) norm was previously introduced to solve several $\ell_0$-constrained optimization problems [44]. In this work, we consider the sparse reconstruction problem by formulating an $\ell_0$-constranined problem. While our solution starts with the same top-($K$,1) norm technique, different from [44], our approach incorporates the classic gradient descent into the DC programming framework to formulate novel sparse recovery algorithms, which are our main contributions.}}, which is simply the sum of $K$ largest-magnitude elements of a vector in terms of absolute values.
	Thus, the original $\ell_0$-minimization sparse reconstruction problem can be equivalently transformed into a least squares optimization penalized by a nonconvex regularizer.
	The nonconvex regularizer is more favorable than the $\ell_1$-regularizer because it removes the penalties on large-\bb{magnitude} elements.
	Although the resulting optimization problem is nonconvex, we can solve it by employing the DC programming framework to transform the problem into a list of convex subproblems, and then we elegantly apply a gradient projection descent method to solve these convex subproblems.
	We propose three different DC gradient projection sparse reconstruction (DC-GPSR) algorithms.
	The proposed algorithms can achieve similar reconstruction accuracies with different time complexities.
	
	The proposed DC-GPSR algorithms have the following advantages in solving a sparse reconstruction problem compared with conventional sparse recovery algorithms.
	First, while conventional algorithms approximate the \mbox{$\ell_0$-constraint} by \mbox{$\ell_1$-relaxation}, we can represent exactly the $\ell_0$-norm constraint using a DC function, i.e., subtracting the top-$(K, 1)$ norm from the $\ell_1$-norm of the reconstructing vector.
	Second, we formulate the sparse recovery as a nonconvex optimization problem whose objective function can be decomposed into DC functions, so that we can naturally adopt the classic DC programming for the remaining optimization tasks.	
	Furthermore, we can perfectly incorporate the gradient projection descent algorithm into the DC programming framework. On one hand, both the gradients of top-$(K, 1)$ norm and $\ell_1$-norm can be easily and quickly calculated. 
	On the other hand, we transform the convex subproblem into a bound-constrained quadratic program (BCQP) with a nonnegativity constraint so that the orthogonal projection operation is simple to compute.
	\bb{Finally, we show that by different DC decompositions and step size strategies we can derive different algorithms that can achieve similar accuracies with different convergence rates. More specifically, the proposed Algorithm 1 has a normal double-loop procedure of the DC programming algorithm, while Algorithm 2 simplifies the double-loop procedure to be a simple single-loop algorithm by using a special DC decomposition. Based on Algorithm 2, which has a fixed step size, we further design new step sizes and propose Algorithm 3 that has a single loop and an accelerated convergence rate.}
	
	The contributions of this paper are summarized as follows:

\begin{itemize}

\item \bb{We propose a double-loop DC-GPSR algorithm (Algorithm 1), which has been reported in our preliminary study [45]. Incorporated by gradient projection descent, the DC programming framework leads to a double-loop DC-GPSR algorithm that shows high accuracy and robustness.} 
\item \bb{We propose a single-loop DC-GPSR algorithm (Algorithm 2), which applies a special DC decomposition to the objective function such that the convex subproblem in the DC programming has a closed-form solution. Compared to Algorithm 1, Algorithm 2 avoids inner loops and solves the DC programming subproblem in a one-step update.} 
\item We make an important observation that an update of the single-loop DC-GPSR algorithm can be interpreted as an update of gradient projection descent with a required step size. Thus, we adopt the Barzilai-Borwein (BB) step size to generalize the basic single-loop DC-GPSR algorithm and propose Algorithm 3 to improve convergence significantly. Compared to Algorithm 1, the runtime of Algorithm 3 is demonstrated to be about $1.5$ times faster in noiseless scenarios and about $20$ times faster in noisy scenarios for wide-range signal-to-noise (SNR) values.
\end{itemize}

\section{System Model}
    We consider a downlink massive MIMO system having $N_t$ transmit antennas at the BS and $N_r$ receiver antennas at the UE.
    Assuming the block-fading narrowband multipath channel \footnote{In this work, we consider narrowband block-fading channels of massive MIMO. It is worth mentioning that our proposed algorithms are generic sparse reconstruction algorithm that can be conveniently applied to other types of sparse channels as long as the channel has sparse representations in certain domains. For example, our proposed algorithms can be applied to channel estimation of wideband time-varying channels for massive MIMO systems. The only required modification is to exploit the channel sparsity in a virtual angular-delay-Doppler domain.}, the channel can be represented by the Saleh-Valenzuela model \footnote{For simplicity, we adopt the basic Saleh-Valenzuela channel model by assuming that the each scattering cluster consists of a single path. Our proposed approach can also be applied to the extended Saleh-Valenzuela channel model, where each cluster consists of multiple sub-paths.} as \cite{tse2005fundamentals} 
    \begin{IEEEeqnarray*}{lCl} 
    \label{spatial_domain_channel}
    \mathbf H_s  = \sqrt{N_r N_t}\sum_{l = 1}^{N_p} \alpha_l \bm{\alpha}_r(\theta_{r, l}) \bm{\alpha}_t^{H}(\theta_{t, l})
	\IEEEyesnumber
    \end{IEEEeqnarray*}
    where $\mathbf H_s \in \mathbb{C}^ {{N_r}\times{N_t}}$ is the channel matrix, $N_p$ is the number of paths, $l$ denotes the multipath index, $\alpha_l$ is the complex channel gain of the $l$th path, $\bm{\alpha}_r(\theta_{r, l})$ and $\bm{\alpha}_t(\theta_{t, l})$ are respectively the array steering vectors of receiver and transmitter array for the $l$th path, and the corresponding angle of arrival (AoA)  and the angle of departure (AoD) are $\theta_{r, l}$ and $\theta_{t, l}$. 
    An array steering vector represents the array phase profile as a function of physical AoA or AoD.
    For a one-dimensional uniform linear array (ULA) consisting of $N$ elements, \bb{the array steering vector can be expressed as $ \bm \alpha (\theta) = \frac{1}{\sqrt{N}} [1, e^{-j2\pi \vartheta}, e^{-j4\pi \vartheta}, ..., e^{-j2\pi \vartheta (N-1)}]^{T}$}, where $\vartheta \in[-1, 1]$ is the normalized spatial angle, which is related to the physical angle $\theta \in [-\pi /2, \pi /2]$ by $\vartheta = \frac{d}{\lambda} \sin{(\theta)}$, and where $d = \lambda/2$ is the antenna spacing and $\lambda$ is the wavelength.

    The channel matrix $\mathbf{H}_s$ can be represented using a two-dimensional Fourier transformation of the angular-domain channel matrix $\mathbf{H}_a$ by \cite{brady2013beamspace} 
    \begin{IEEEeqnarray*}{lCl}
    \label{beamspace_spatial_transformation}
   \mathbf H_s &=& \mathbf U_r \mathbf H_a \mathbf U_t^H \\
		 	   &=& \sum_{i=0}^{N_r-1} \sum_{j=0}^{N_t-1} H_a (i, j) \bm{\alpha}_r(\theta_{r,i}) \bm \alpha_t^H(\theta_{t,j})
    \IEEEyesnumber
    \end{IEEEeqnarray*}
    \bb{where $\mathbf{U}_t \in \mathbb{C}^{N_t \times N_t}$ and $\mathbf{U}_r \in \mathbb{C}^{N_r \times N_r}$ are unitary digital Fourier transform (DFT) matrices, and they can be expressed using array steering vectors as $\mathbf{U}_t = [\bm{\alpha}_t(\theta_{t,0}), \bm{\alpha}_t(\theta_{t,1}),...,\bm{\alpha}_t(\theta_{t, N_t-1})]^{T}$, $\mathbf{U}_r = [\bm{\alpha}_r(\theta_{r,0}), \bm{\alpha}_r(\theta_{r,1}),...,\bm{\alpha}_r(\theta_{r, N_r-1})]^{T}$}, where $\{\theta_{t,0}, ..., \theta_{t,N_t-1}\}$ and $\{\theta_{r,0}, ..., \theta_{r,N_r-1}\}$ are the virtual AoDs and AoAs defined by array elements of transmitter and receiver. The beamspace channel $\mathbf H_a$ is sparse due to the limited number of multiple paths and the high dimensionality of massive MIMO channels \cite{tse2005fundamentals}.
    	
	By transmitting the known pilots $\mathbf{P} \in \mathbb{C}^{N_t \times L}$ from the BS through the channel, we have the pilot signal received at UE as
    \begin{equation} \label{received_signal}
    \tilde{\mathbf{R}} = \mathbf H_s \mathbf{P} + \tilde{\mathbf{W}} 
    \end{equation}
    where $\tilde{\mathbf{R}} \in \mathbb{C}^{N_r \times L}$ is the received pilot observations, and where $L$ is the length of training pilot sequence for each antenna; $\mathbf H_s \in \mathbb{C} ^ {N_r \times N_t}$ is the channel matrix in the spatial domain, $\mathbf{P}$ is the transmitted pilot matrix; $\tilde{\mathbf{W}} \in \mathbb{C}^{N_r \times L}$ is the additive white Gaussian noise matrix whose elements are independent identical distributed (i.i.d.) complex Gaussian random variables having a mean of zero and a variance of $\sigma_n^2$.
	Conventional estimation methods such as the linear minimum mean square error (LMMSE) or the least squares (LS) require $L \geq N_t$ to obtain accurate estimates, whereas the beamspace channel estimation can reconstruct $\hat{\mathbf{H}}_s$ for $L \ll N_t$.	
    From \eqref{beamspace_spatial_transformation}, we have $\mathbf H_s = \mathbf{U}_r\mathbf{H}_a \mathbf{U}_t^{H}$, which can be substituted in \eqref{received_signal} to have
    \begin{IEEEeqnarray*}{lCl}
    \label{received signal in beamspace}
    \tilde{\mathbf{R}} =  \mathbf{U}_r \mathbf{H}_a \mathbf{U}_t^{H} \mathbf{P} + \tilde{\mathbf{W}}.
    \IEEEyesnumber
    \end{IEEEeqnarray*}
    By taking the transposes and right multiplications with $\mathbf{U}_r^H$ on both sides of \eqref{received signal in beamspace}, we have
    \begin{IEEEeqnarray*}{lCl}
    \label{conjugate transform multi antenna}
    \underbrace{\tilde{\mathbf{R}}^T \mathbf{U}_r^{H}}_{\mathbf {R}} 
    &=  
    \underbrace{\mathbf{P}^T \mathbf{U}_t^{*}}_{\mathbf S} 
    \underbrace{\mathbf{H}_a^T \mathbf{U}_r \mathbf{U}_r^{H}}_{\mathbf H}  
    + \underbrace{\tilde{\mathbf{W}}^T \mathbf{U}_r^{H}}_{\mathbf W}
    \IEEEyesnumber
    \end{IEEEeqnarray*}
	We simply write \eqref{conjugate transform multi antenna} as 
     \begin{IEEEeqnarray*}{lCl} \label{CS_multi_antenna}
    \mathbf R =\mathbf S \mathbf{H} + \mathbf W   
     \IEEEyesnumber  
    \end{IEEEeqnarray*}
    where $\mathbf {R}= \tilde{\mathbf R}^T \mathbf{U}_r^{H} \in \mathbb{C}^{L \times N_r}$, $\mathbf{H} = \mathbf{H}_a^T \mathbf{U}_r \mathbf{U}_r^{H} \in \mathbb{C}^{N_t \times N_r}$, $\mathbf W \in \mathbb{C}^{L \times N_r}$, and the matrix $\mathbf S = \mathbf{P}^T \mathbf{U}_t^* \in \mathbb{C}^{L \times N_t}$. 
	The beamspace channel estimation $\hat{\mathbf{H}}=\hat{\mathbf{H}}_a^T$ can be obtained by solving the linear equation \eqref{CS_multi_antenna} given the known measurement matrix $\mathbf{S}$ and measurements $\mathbf{R}$.
	To this end, both an efficient sparse recovery algorithm and a proper measurement matrix are essential to the estimation performances. 
	Eq. \eqref{conjugate transform multi antenna} suggests that the pilot matrix $\mathbf P$ depends on the measurement matrix $\mathbf{S}$.
	To design the pilot matrix, we can first determine a measurement matrix $\mathbf{S}$, and then obtain the pilot matrix by $\mathbf P = \mathbf{S}^T \mathbf{U}_t$, where $ \mathbf{U}_t$ is a DFT matrix.
	Due to $\norm{\mathbf S}_F^2 = \norm{\mathbf{P}^T \mathbf{U}_t^*}_F^2$, the power constraint on a pilot matrix $\norm{\mathbf P}_F^2 = P$ can be imposed via scaling the measurement matrix by $\mathbf S = P \frac{\mathbf S^\prime}{\norm{\mathbf S^\prime}_F^2}$, where $\norm{\cdot}_F$ represents the Frobenius norm of a matrix.
	Various random matrices can be adopted as the measurement matrix.
	After evaluating the proposed reconstruction algorithms that adopt various random matrices (such as Gaussian, Bernoulli, and partial Fourier matrices), we conclude that these random matrices have similar reconstruction performances.

	\bb{To solve the multiple measurement vector (MMV) problem \eqref{CS_multi_antenna}, we propose the column-wise broadcasting vectorization method to vectorize the channel matrix $\mathbf H$. Thus, the MMV problem \eqref{CS_multi_antenna} can be transformed into parallel single measurement vector (SMV) problems. Therefore, although the proposed algorithms are developed for an SMV model, they can be extended to the MMV problem.}	
	We express the linear equation \eqref{CS_multi_antenna} as 
    \begin{IEEEeqnarray*}{lCl}
    [\mathbf r_1,\mathbf r_2,..., \mathbf r_{N_r}]
				&= \mathbf S [\mathbf h_1,\mathbf h_2,..., \mathbf h_{N_r}]
				+ [\mathbf w_1,\mathbf w_2,..., \mathbf w_{N_r}]
	\IEEEyesnumber
    \end{IEEEeqnarray*}
	where $\mathbf r_i$ is the $i$th ($1 \le i \le N_r$) column of $\mathbf{R}$; $\mathbf h_i$ is the $i$th ($1 \le i \le N_r$) column of $\mathbf{H}$; $\mathbf w_i$ is the $i$th ($1 \le i \le N_r$) column of $\mathbf W$.
    For the $i$th column of $\mathbf{R}, \mathbf{H}$ and $\mathbf W$ in \eqref{CS_multi_antenna}, we have 
    \begin{IEEEeqnarray*}{lCl} \label{CS_formulation}
    \mathbf r_i = \mathbf S \mathbf h_i + \mathbf w_i.
    \IEEEyesnumber
    \end{IEEEeqnarray*}
	We can further express \eqref{CS_formulation} by an equivalent real-valued linear equation as
	\begin{IEEEeqnarray*}{lCl} \label{real_sepa_imag}
	\underbrace{\left[ \begin{array}{cc}
		\Re(\mathbf r_i) \\ \Im(\mathbf r_i)
	\end{array} \right]}_{\mathbf y}
	= 
	\underbrace{\begin{bmatrix}
	\Re(\mathbf S) & -\Im(\mathbf S) \\
	\Im(\mathbf S) & \Re(\mathbf S)
	\end{bmatrix}}_{\mathbf \Phi}
	\underbrace{
	\begin{bmatrix}
	\Re(\mathbf h_i) \\ \Im(\mathbf h_i)
	\end{bmatrix}}_{\mathbf x}
	+ 	
	\underbrace{\begin{bmatrix}
	\Re(\mathbf w_i) \\ \Im(\mathbf w_i)
	\end{bmatrix}}_{\mathbf n}
	\IEEEyesnumber.
	\end{IEEEeqnarray*} 
	Thus, we obtain an SMV problem \eqref{real_sepa_imag}.
	For the presentational simplicity, we concisely express \eqref{real_sepa_imag} as 
	\begin{IEEEeqnarray*}{lCl} \label{CS_real}
    		\mathbf y = \mathbf \Phi \mathbf x + \mathbf n.
    		\IEEEyesnumber
	\end{IEEEeqnarray*}
	In the remainder of this paper, we uniquely refer the real-form vector $\mathbf x = [\Re(\mathbf h_i)^T, \Im(\mathbf h_i)^T]^T$ in \eqref{CS_real} as the equivalent sparse channel vector, since the vector $\mathbf x$ contains the equivalent information of a complex beamspace channel $\mathbf{h}_i$. We omit the subscript $i$ since \eqref{CS_real} is valid for arbitrary beamspace channel $\mathbf{h}_i$ for $1 \le i \le N_r$.

\section{DC Representation for Sparse Constraint}
	In this section, \bb{we} first briefly review the sparse reconstruction optimization problem. 
	Then, we introduce the \mbox{top-$(K,1)$} norm \cite{Gotoh2018DC} to represent exactly the $\ell_0$-norm constraint in the sparse reconstruction optimization problem. Also, we provide a threshold to determine the range of the penalty parameter. 
	
	According to \eqref{CS_real}, reconstructing $\mathbf x$ from $\mathbf y$ and $\mathbf \Phi$ using the sparsity as a priori is an NP-hard \mbox{$\ell_0$-minimization} problem, which is defined as \cite{sparse2015rish}
	\begin{IEEEeqnarray*}{lCl}
	\label{sparse_recovery1}
	\mathop{\text{min}}\limits_{\mathbf{x}} &&\quad \norm{\mathbf{x}}_0 \\
	\text{s.t.} &&\quad \norm{\mathbf{y}-\mathbf{\Phi}\mathbf{x}}_2^2 \le \tau
	\IEEEyesnumber
	\end{IEEEeqnarray*} 
	where $\tau$ is nonnegative and real.
	Problem \eqref{sparse_recovery1} can be rewritten in an equivalent form as \cite{sparse2015rish}
	\begin{IEEEeqnarray*}{lCl}
	\label{sparse_recovery2}
	\mathop{\text{min}}\limits_{\mathbf{x}} &&\quad  \norm{\mathbf{y}-\mathbf{\Phi}\mathbf{x}}_2^2\\
	\text{s.t.} &&\quad \norm{\mathbf{x}}_0  \le K
	\IEEEyesnumber
	\end{IEEEeqnarray*} 
	where $K$ is an upper bound for the number of nonzero elements in $\mathbf x$, and it is uniquely determined by the parameter $\tau$ in \eqref{sparse_recovery1}.

	Instead of using the common $\ell_1$-relaxation, we introduce the top-$(K,1)$ norm to seek an equivalent expression for the constraint $\norm{\mathbf{x}}_0  \le K$ in the original problem \eqref{sparse_recovery2}.
	 The top-$(K,1)$ norm $\norm{\mathbf x}_{K,1}$ is defined as the sum of the largest $K$ elements of the vector $\mathbf x$ in terms of the absolute values, namely 
	\begin{IEEEeqnarray*}{lCl}
	\norm{\mathbf x}_{K,1} :=  | x_{(1)} | +| x_{(2)} | + \cdots +| x_{(K)} |
	\IEEEyesnumber
	\end{IEEEeqnarray*} 
	where $| x_{(i)}|$ denotes the element whose absolute value is the $i$th-largest among the $N$ elements of the vector $\mathbf x$, i.e., $ | x_{(1)} | \ge| x_{(2)} | \ge \cdots \ge | x_{(N)} |$. The constraint $\norm{\mathbf{x}}_0  \le K$ is equivalent to the statement that the $(K+1)$th-largest element of the vector $\mathbf x$ is zero, i.e., $\norm{\mathbf x}_{K+1, 1} - \norm{\mathbf x}_{K,1} = 0$. 
	Thus, we have an equivalent relationship between the following two statements \cite{Gotoh2018DC} 
	\begin{IEEEeqnarray*}{lCl}
	\label{equivalence}
	\norm{\mathbf{x}}_0  \le K \Leftrightarrow \norm{\mathbf x}_1 - \norm{\mathbf x}_{K,1}=0.
	\IEEEyesnumber
	\end{IEEEeqnarray*} 
	Since both $\norm{\mathbf x}_1$ and $\norm{\mathbf x}_{K,1}$ are convex, we say the equality $\norm{\mathbf x}_1 - \norm{\mathbf x}_{K,1}= 0$ is an exact DC representation for the sparsity constraint.
	By replacing the sparsity constraint $\norm{\mathbf x}_0 \le K$ in \eqref{sparse_recovery2} using the DC constraint $\norm{\mathbf x}_1 - \norm{\mathbf x}_{K,1}= 0$, we rewrite the sparse reconstruction problem as 
	\begin{IEEEeqnarray*}{lCl}
	\label{sparse_recovery3}
	\mathop{\text{min}}\limits_{\mathbf{x}} 
	&&\quad  \norm{\mathbf{y}-\mathbf{\Phi}\mathbf{x}}_2^2 \\
	\text{s.t.} 
	&&\quad \norm{\mathbf x}_1 - \norm{\mathbf x}_{K,1}=0.
	\IEEEyesnumber
	\end{IEEEeqnarray*} 
	Using an appropriate Lagrange multiplier $\rho$, from \eqref{sparse_recovery3} we obtain the following unconstrained optimization problem  
	 \begin{IEEEeqnarray*}{lCl}	
	 \label{dc_penalty_function}
	\mathop{\text{min}}\limits_{\mathbf{x}} \quad 
	\frac{1}{2}\norm{\mathbf{y}-\mathbf{\Phi}\mathbf{x}}_2^2 
	+ \rho (\norm{\mathbf x}_1 - \norm{\mathbf x}_{K,1})
	:=	F(\mathbf x) 
	\IEEEyesnumber
	\end{IEEEeqnarray*} 
	where $\rho$ is the regularization parameter that balances the data consistency and the penalty term.
	Our formulated optimization problem \eqref{dc_penalty_function} differs from the conventional $\ell_1$-regularized sparse reconstruction \footnote{$\mathop{\text{min}}\limits_{\mathbf x}	\frac{1}{2}\norm{\mathbf y - \mathbf \Phi \mathbf x}_2^2 + \lambda \norm{\mathbf x}_1$} only in terms of the subtracted \mbox{top-$(K, 1)$} norm $\norm{\mathbf x}_{K,1}$ in its penalty term.
	The regularizer $\rho (\norm{\mathbf x}_1 - \norm{\mathbf x}_{K,1})$ is better than an $\ell_1$-norm regularizer because it removes the penalties on the $K$ largest-magnitude elements \footnote{In practice, if the sparsity level of reconstructing vector is already known, we set $K$ as the number of nonzero elements of reconstruction vectors; otherwise, we can use cross validation to determine an approximation of $K$.}.
	
	To ensure the equivalence between the unconstrained problem \eqref{dc_penalty_function} and the constrained problem \eqref{sparse_recovery3}, the following theorem specifies the range for the penalty parameter. 
	
	\begin{theorem}
	Let $\mathbf x_{\rho^*}$ be an optimal solution to \eqref{dc_penalty_function} with given $\rho^*$.
	Suppose there exists a constant $ q > 0$ such that $\norm{\mathbf x_{\rho^*}}_2 \le q$ for any $\rho^* >0$.
	Then $\mathbf x_{\rho^*}$ is also optimal to \eqref{sparse_recovery3} if 
	$$
	\rho^*\ge \mathop{\text{max}}_{i}\{q(\norm{\mathbf \Phi ^T \mathbf \Phi \mathbf e_i}_2 + |(\mathbf{\Phi}^T\mathbf{\Phi})_{ii}|/2 )+|(\mathbf \Phi^T \mathbf y)_{i}| \}
	$$
	where $1 \le i\le N_t$; $\mathbf e_i$ represents the unit vector in which the $i$th element is one while the other elements are zeros; $(\mathbf{\Phi}^T\mathbf{\Phi})_{ii}$ represents the $i$th diagonal elements of matrix $\mathbf \Phi^T \mathbf \Phi$; $(\mathbf \Phi^T \mathbf y)_i$ indicates the $i$th element of the vector $\mathbf \Phi^T \mathbf y$.
 	\end{theorem}
	\emph{Proof:} See Appendix A.
	
	Theorem 1 indicates that given a penalty parameter $\rho$ having a suitably large value, the optimal solution to \eqref{dc_penalty_function} is also the optimal solution to \eqref{sparse_recovery3}. 
	To calculate the lower threshold $\rho^*$, we can first estimate a constant $q$ such that $\norm{\mathbf x_{\rho^*}}_2 \le q$. 
	In practice, to avoid high \bb{computational} complexity of the inequality in Theorem 1, we can use the cross validation to select a suitable value for the penalty parameter $\rho$.

\section{Double-Loop DC Gradient Projection Descent for Sparse Reconstruction}
 	We have formulated the sparse reconstructions into a nonconvex optimization problem \eqref{dc_penalty_function}.
 	In this section, we use DC programming and gradient projection descent to solve \eqref{dc_penalty_function} and propose a double-loop DC-GPSR algorithm.

\subsection{DC Programming Framework}
	For a nonconvex unconstrained optimization problem
	\begin{IEEEeqnarray*}{lCl}
	\label{dc_general_obj}
	\mathop {\text{min}}_\mathbf x 
	f(\mathbf x) - g(\mathbf x)
	\IEEEyesnumber
	\end{IEEEeqnarray*}	 
	where $f(\mathbf x)$ and $g(\mathbf x)$ are two convex functions. The DC programming method solves the following convex subproblem at the $t$th-iteration,
	\begin{IEEEeqnarray*}{lCl}
	\label{general_dc_sub}
	\mathop{\text{min}}\limits_{\mathbf{x}}  f(\mathbf x)
	- \mathbf x^T \partial g(\mathbf x^{t-1})
	\IEEEyesnumber
	\end{IEEEeqnarray*}
	where the second convex function $g(\mathbf x)$ in \eqref{dc_general_obj} is linearized by $\mathbf x^T \partial g(\mathbf x^{t-1})$ in \eqref{general_dc_sub}, and where $\partial g(\mathbf x^{t-1})$ represents the gradient (or subgradient) of $g(\mathbf x^{t-1})$ with respect to $\mathbf x^{t-1}$.
	The DC algorithm framework can be outlined as follows:
	\begin{itemize}
	\item[1.] \textbf{Start:} Given a starting point $\mathbf x ^0$, and a terminate condition
	\item[2.] \textbf{Repeat:} For $t=1,2,...$ \\ 
					\hspace*{4mm} Compute the gradient (or subgradient) $ \partial g(\mathbf x^{t-1})$. \\
					\hspace*{4mm} Solve the convex subproblem \eqref{general_dc_sub} to obtain $\mathbf x^t$.
	\item[3.] \textbf{End:} Until a terminate condition is satisfied.				
	\end{itemize}
	
	The DC programming is an iterative algorithm framework that can ensure global convergence, which means the DC algorithm can converge from an arbitrary initial point \footnote{The global convergence property has been comprehensively studied for general DC algorithms, hence it is also valid for the proposed algorithms. For the convergence properties and proofs of DC algorithms, we refer the interested readers to \cite[Sec. 3.2]{Tao1997Convex}, the \cite[Sec. 2]{Sriperumbudur2912A}, \cite[p. 10]{Thi2018DC}, \cite[pp. 5-8]{Dinh2014Recent} and the references therein.}.

\subsection{A Double-Loop DC-GPSR Algorithm}
	Following the DC programming framework, we decompose our objective function in \eqref{dc_penalty_function} as the difference of the two convex functions of $f(\mathbf x)$ and $g(\mathbf x)$ 
	\begin{IEEEeqnarray*}{lCl}
	\label{dc_quadratic_proj}
	\mathop {\text{min}}_\mathbf x 
	\underbrace{ \frac{1}{2}\norm{\mathbf{y}-\mathbf{\Phi}\mathbf{x}}_2^2  + \rho \norm {\mathbf x}_1}_{f(\mathbf x)}
	- \underbrace{ \rho \norm {\mathbf x}_{K,1}}_{g(\mathbf x)}.
	\IEEEyesnumber
	\end{IEEEeqnarray*}	 
	At the $t$th-iteration, we solve the following convex subproblem 
	\begin{IEEEeqnarray*}{lCl}
	\label{2loop_general_dc}
	\mathop{\text{min}}\limits_{\mathbf{x}} \quad  f(\mathbf x)
	- \rho \mathbf x^T \partial \norm{\mathbf x^{t-1}}_{K,1}
	\IEEEyesnumber
	\end{IEEEeqnarray*}
	 where $\partial \norm{\mathbf x^{t-1}}_{K,1}$ denotes the subgradient of \mbox{top-$(K,1)$} norm of $\mathbf x^{t-1}$, and where the superscript $t-1$ indicates the $(t-1)$th update.
	The subgradient $\partial \norm{\mathbf x}_{K,1}$ of the \mbox{top-$(K,1)$} norm of $\mathbf x$ is defined as \cite{Gotoh2018DC}
	\begin{IEEEeqnarray*}{lCl}	
	\label{2loop_general_dc_sub}
	 \partial \norm{\mathbf x}_{K,1}
	 := \mathop{\text{argmax}}_\mathbf w \left\lbrace \sum_{i=1} ^{N} x_i w_i \Big\vert \sum_{i=1}^N |w_i| =K,  w_i \in [-1, 1]\right\rbrace.\\
	\IEEEyesnumber
	\end{IEEEeqnarray*} 
	By substituting the $f(\mathbf x)$ defined in \eqref{dc_quadratic_proj} into \eqref{2loop_general_dc} and denoting $\mathbf w_x^{t-1}$ by a feasible value for the subgradient $\partial \norm{\mathbf x^{t-1}}_{K,1}$, we write the subproblem \eqref{2loop_general_dc} as 
	\begin{IEEEeqnarray*}{lCl}
	\label{subprob_quadratic_proj}
	\mathop{\text{min}}\limits_{\mathbf{x}} \quad 
	\frac{1}{2}\norm{\mathbf{y}-\mathbf{\Phi}\mathbf{x}}_2^2
	+ \rho \norm{\mathbf x}_1 
	- \rho \mathbf x^T \mathbf w_x^{t-1}
	\IEEEyesnumber
	\end{IEEEeqnarray*}
	where $\mathbf w_x^{t-1} \in \partial \norm{\mathbf x^{t-1}}_{K,1}$. A feasible subgradient $\mathbf w_x^{t-1}$ can be simply obtained by setting the signs of the first $K$ largest elements of $|\mathbf{x}^{t-1}|$ to the corresponding elements of $\mathbf w_x^{t-1}$, i.e., $({\mathbf w_{x}})_{(i)}^{t-1}=\sign(x_{(i)}^{t-1})$, where the subscript $i$ indicates the $i$th element of a vector, and setting the other elements of $\mathbf w_x^{t-1}$ to be zeros.
	
	We obtain a convex subproblem \eqref{subprob_quadratic_proj}.
	We will turn \eqref{subprob_quadratic_proj} into a constrained quadratic problem so that we can solve it using the gradient projection descent method.
	Specifically, we split the positive and negative part of $\mathbf x$, and represent $\mathbf x$ as the difference of its positive part $\mathbf u$ and its negative part $\mathbf v$, that is
	\begin{IEEEeqnarray*}{lCl}
	\label{pos_neg_x}
	\mathbf x = \mathbf u - \mathbf v, \quad \mathbf u \succeq \mathbf 0, \mathbf v \succeq \mathbf 0.
	\IEEEyesnumber
	\end{IEEEeqnarray*}
	where $\mathbf u = (\mathbf x)_+, \mathbf v = (-\mathbf x)_+$, and where $(\cdot)_+$ is the positive-taking operation that retains the positive elements and sets the other elements to be zeros. More precisely, $(\mathbf x)_+$ represents for each element $x$ in vector $\mathbf x$ we take $(x)_+ = \max \{0, x\}$;  $(-\mathbf x)_+$ represents for each element $-x$ in vector $-\mathbf x$ we take $(-x)_+ = \max \{0, -x\}$.
	Noticing that $\norm{\mathbf x}_1 = \mathbf 1^T \mathbf u 	+  \mathbf 1^T \mathbf v$,
	the subproblem \eqref{subprob_quadratic_proj} can be written as a BCQP problem
	\begin{IEEEeqnarray*}{lCl}
	\label{bcqp}
	 \mathop{\text{min}}\limits_{\mathbf{u, v}} \quad 
	 && \frac{1}{2}\norm{\mathbf{y}-\mathbf{\Phi}(\mathbf{u} - \mathbf v)}_2^2 
	 	+ \rho \mathbf 1^T \mathbf u  	+ \rho \mathbf 1^T \mathbf v \\
	 && - \rho \mathbf u^T \mathbf w_u^{t-1} - \rho \mathbf v^T \mathbf w_v^{t-1} \\
	  \text{s.t.} \quad && \mathbf{u} \succeq \mathbf{0}, \mathbf{v} \succeq \mathbf{0}
	\IEEEyesnumber
	\end{IEEEeqnarray*}
	where $\mathbf w_u^{t-1}$ and $\mathbf w_v^{t-1}$, respectively, represent the positive and negative part of $\mathbf w_x^{t-1}$, i.e., $\mathbf w_u^{t-1} = (\mathbf w_x^{t-1})_+$, $\mathbf w_v^{t-1} = (-\mathbf w_x^{t-1})_+$.
	Let $\mathbf{z}$ denote the concatenation of $\mathbf{u}$ and $\mathbf{v}$, i.e., $\mathbf{z} = [\mathbf{u}^{T}, \mathbf{v}^{T}]^{T}$,
	we rewrite \eqref{bcqp} into a compact form
\begin{IEEEeqnarray*}{lCl}\label{compact_bcqp}
    \mathop{\text{min}}\limits_{\mathbf{z}} \quad 
    &&  \frac{1}{2} \mathbf{z}^{T} \mathbf{B} \mathbf{z} 
    + \mathbf c^T \mathbf z
    := G(\mathbf{z}) ,\\ 
    \text{s.t.} \quad && \mathbf{z} \succeq \mathbf{0}
    \IEEEyesnumber
\end{IEEEeqnarray*}
where 
    \begin{IEEEeqnarray*}{lCl}
     \mathbf{z} = \left[\begin{matrix}
    \mathbf{u}\\
    \mathbf{v}
    \end{matrix}\right],
    \quad
    \mathbf{B} = \left[
    \begin{matrix}
    \mathbf{\Phi}^{T} \mathbf{\Phi} & -\mathbf{\Phi}^{T} \mathbf{\Phi}\\
    - \mathbf{\Phi}^{T} \mathbf{\Phi} & \mathbf{\Phi}^{T} \mathbf{\Phi}
    \end{matrix}
    \right],
    \quad \\
    \mathbf c = \left[\begin{matrix}
    -\mathbf{\Phi}^{T} \mathbf{y}\\
    \mathbf{\Phi}^{T} \mathbf{y}
    \end{matrix}\right]
    +\rho \mathbf 1^T
    - \rho \mathbf w_z^{t-1} 
	\end{IEEEeqnarray*}
where $\mathbf 1^T$ represents an all-ones column vector having the same dimension as $\mathbf z$, and $\mathbf w_z^{t-1} = [(\mathbf w_u^{t-1})^T, (\mathbf w_v^{t-1})^T]^T$.
Note that $\mathbf w_z^{t-1} $ is a subgradient of $\norm{\mathbf z^{t-1}}_{K,1}$, i.e., $ \mathbf w_z^{t-1} \in \partial \norm{\mathbf z^{t-1}}_{K,1}$. 
	Since $\mathbf z^{t-1} \succeq \mathbf 0$, a feasible subgradient $\mathbf w_z^{t-1} $ can be an indicator vector having either one-valued or zero-valued elements, where the indices for the one-valued elements of $ \mathbf w_z^{t-1}$ correspond to the indices of the $K$-largest elements of $\mathbf z^{t-1}$.  	
	Eq. \eqref{compact_bcqp} is an equivalent problem for the subproblem \eqref{subprob_quadratic_proj}.
	We apply the gradient projection descent method to solve \eqref{compact_bcqp}, thus the $k$th update is 
    \begin{IEEEeqnarray*}{lCl}
    \label{accelerated_pgd_general}
    \mathbf{z}^{(k+\frac{1}{2})} &= \textit{Proj} \left(\mathbf{z}^{(k)} - \alpha^k \nabla G(\mathbf{z}^{(k)})\right),\\
     \mathbf{z}^{(k+1)} &= \mathbf{z}^{(k)} +\beta^{k} ( \mathbf{z}^{(k+\frac{1}{2})}  -  \mathbf{z}^{(k)} )
     \IEEEyesnumber
    \end{IEEEeqnarray*}
    where $\alpha^{k} > 0$ is the step size and it can be determined by the BB step size, which can be calculated as $\alpha^{t} = \frac{\norm{\mathbf z^{k}- \mathbf z^{k-1}}^2}{(\mathbf z^{k} - \mathbf z^{k-1})^T \left( G(\mathbf z^k) - G(\mathbf z^{k-1}) \right)}$; $\beta^{k} \in (0, 1]$ is another step size to ensure the monotonic-decrease of the objective and it can be calculated in closed-form as $\beta^{k} =\frac{(\bm \delta^{k})^T \nabla G(\mathbf z^k)}{(\bm \delta^{k})^T \mathbf B \bm \delta^k}$, where $\bm \delta^{k}=\mathbf{z}^{(k+\frac{1}{2})}  -  \mathbf{z}^{(k)}$ \cite{figueiredo2007gradient};
    $\textit{Proj}(\cdot)$ represents the operation of orthogonal projection that projects the vector onto the nonnegative orthant \footnote{Let $\mathbb{R}_+^n=\{ \mathbf x=(x_1, x_2, ..., x_n) | x_1\ge 0, x_2\ge 0,..., x_n\ge 0 \}$ be the nonnegative orthant of $\mathbb{R}^n$.};
     $\nabla G(\mathbf{z}^{(k)})$ represents the gradient of $G(\mathbf z)$ defined in \eqref{compact_bcqp} with respect to $\mathbf{z}^{(k)}$. We have $\nabla G(\mathbf{z}^{(k)})=\mathbf B \mathbf z + \mathbf c$, which can be calculated by
    \begin{IEEEeqnarray*}{lCl}
    \label{gradient_Gz}
   \nabla G(\mathbf{z}^{(k)}) 
  	=  
   && \left[
    \begin{matrix}
    \mathbf{\Phi}^{T} \mathbf{\Phi} (\mathbf{u}^{(k)}-\mathbf{v}^{(k)})\\
    -\mathbf{\Phi}^{T} \mathbf{\Phi} (\mathbf{u}^{(k)}-\mathbf{v}^{(k)})
     \end{matrix}
    \right]
    + \left[\begin{matrix}
    -\mathbf{\Phi}^{T} \mathbf{y}\\
    \mathbf{\Phi}^{T} \mathbf{y}
    \end{matrix}\right] \\
    && - \rho \mathbf w_z^{t-1} +\rho \mathbf{1}^T. 
    \IEEEyesnumber
    \end{IEEEeqnarray*}

	In a nutshell, the proposed algorithm computes the following two steps iteratively until convergence:
	\begin{IEEEeqnarray*}{lCl}
	\label{dc_projection}
	(\text{a}) 
	&& \quad \mathbf w_z^{t-1} 
     \in  \partial \norm{\mathbf z^{t-1}}_{K, 1}\\
	(\text{b}) 
	&& \quad \mathbf z^{t} 
	= \mathop{\text{argmin}}_{\mathbf z \succeq 0} 
	\lbrace 
	\frac{1}{2} \mathbf{z}^{T} \mathbf{B} \mathbf{z} + \mathbf c^T \mathbf z	
    \rbrace 
	\IEEEyesnumber  
	\end{IEEEeqnarray*} 
    where $\mathbf z$, $\mathbf{B}$ and $\mathbf c$ are defined in \eqref{compact_bcqp}.
    The step (b) in \eqref{dc_projection} is calculated by applying the gradient projection descent updates in \eqref{accelerated_pgd_general}. We summarize this double-loop DC-GPSR algorithm in Algorithm 1.

\begin{algorithm}
	\caption{Double-loop DC-GPSR (DlDC-GPSR)} 
	\hspace*{\algorithmicindent} \textbf{Input:} measurements $\mathbf y$, measurement matrix $\mathbf \Phi$ and a small number $\epsilon$ \\
    \hspace*{\algorithmicindent} \textbf{Output:} reconstruction $\hat{\mathbf x}$ \\
    	\hspace*{\algorithmicindent} \textbf{Initialization:}  $\mathbf u^0$, $\mathbf v^0$, $\mathbf z^0 \leftarrow [(\mathbf u^0)^T,  (\mathbf v^0)^T]^T$
	\begin{algorithmic}[1]
		\For {$t=1,2,\ldots$}
			\State Compute a subgradient $\mathbf w_z^{t-1} \in \partial \norm{\mathbf z^{t-1}}_{K,1}$. 
			\For {$k=1,2,\ldots$}
				\State Compute gradient $\nabla G(\mathbf{z}^{(k)})$ by \eqref{gradient_Gz}.
				\State Perform gradient projection descent \eqref{accelerated_pgd_general} and obtain $\mathbf z^{(k+1)}$.
				\State Check convergence, set $\mathbf z^* \leftarrow \mathbf z^{(k+1)}$ and proceed to Step 7 if convergence is satisfied; otherwise return to Step 3.
			\EndFor
			\State $\mathbf z^t \leftarrow \mathbf z^*$
			\State Check terminate condition $\norm{\mathbf z^{t}-\mathbf z^{t-1}}_2 \le \epsilon$ and return to Step 1 if not satisfied; otherwise, terminate with $\mathbf z^t =  [(\mathbf u^t)^T,  (\mathbf v^t)^T]^T$, and obtain the reconstruction $\hat{\mathbf x} = \mathbf u^t - \mathbf v^t$.
		\EndFor
	\end{algorithmic} 
\end{algorithm}

\section{Single-Loop DC Gradient Projection Descent for Sparse Reconstruction}
	In the proposed DlDC-GPSR algorithm, at each iteration we solve a nonsmooth convex subproblem \eqref{subprob_quadratic_proj} using another iterative algorithm, i.e., the gradient projection descent updates. 
	This double-loop procedure can be computationally inefficient.
	In this section, we derive a closed-form solution to the convex subproblem by performing a special DC decomposition to eliminate the inner iterations.
Thus, we propose a basic single-loop DC-GPSR algorithm.
	Interestingly, we observe that the proposed basic \mbox{single-loop} DC-GPSR algorithm can be interpreted as simple gradient projection descent updates with a required step size. 
	Furthermore, we accelerate the single-loop \mbox{DC-GPSR} algorithm using the BB step size and propose an extension algorithm for the single-loop \mbox{DC-GPSR}.

\subsection{A Basic Single-Loop DC-GPSR Algorithm}

	 We first rewrite the least squares objective of problem \eqref{dc_penalty_function} in an equivalent form as
	\begin{IEEEeqnarray*}{lCl}
	\label{dc_ls_obj}
	\frac{1}{2}\norm{\mathbf y - \mathbf {\Phi x}}_2^2 = \frac{l}{2} \norm{\mathbf x}_2^2 -
	\left( \frac{l}{2} \norm{\mathbf x}_2^2-\frac{1}{2}\norm{\mathbf y - \mathbf {\Phi x}}_2^2 \right)
		\IEEEyesnumber
	\end{IEEEeqnarray*}
	where $l \ge 0$ is a Lipschitz constant of the least square objective.
	By substituting \eqref{dc_ls_obj} into \eqref{dc_penalty_function}, we write the unconstrained sparse reconstruction problem as
	\begin{IEEEeqnarray*}{lCl}
	\label{obj_proj_close}
	\mathop {\text{min}}_\mathbf x 
	 && \frac{l}{2} \norm{\mathbf x}_2^2 - \left( \frac{l}{2} \norm{\mathbf x}_2^2- \frac{1}{2}\norm{\mathbf y - \mathbf {\Phi x}}_2^2 \right) 
	 + \rho (\norm {\mathbf x}_1 - \norm{\mathbf x}_{K,1})
    \\ &&  :=F(\mathbf x).
	\IEEEyesnumber
	\end{IEEEeqnarray*}
	Then, we perform the following DC decomposition on the objective $F(\mathbf x)$ defined in \eqref{obj_proj_close}, and we have
	\begin{IEEEeqnarray*}{lCl}
	\label{obj_proj_close_dc}
	\mathop {\text{min}}_\mathbf x 
	\underbrace{ \frac{l}{2} \norm{\mathbf x}_2^2 + \rho \norm {\mathbf x}_1}_{f(\mathbf x)}
	 - \underbrace{ \left( \frac{l}{2} \norm{\mathbf x}_2^2- \frac{1}{2}\norm{\mathbf y - \mathbf {\Phi x}}_2^2
	 + \rho \norm{\mathbf x}_{K,1} \right)}_{g(\mathbf x)} \\
	\IEEEyesnumber
	\end{IEEEeqnarray*}
	where the functions $f(\mathbf x)$ and $g(\mathbf x)$ are convex. 
	The convexity of $g(\mathbf x)$ can be ensured by confirming the convexity of $\frac{l}{2} \norm{\mathbf x}_2^2- \frac{1}{2}\norm{\mathbf y - \mathbf {\Phi x}}_2^2$, which is given in Theorem 2.
	
	\begin{theorem}
	The least squares objective $\frac{1}{2}\norm{\mathbf y - \mathbf {\Phi x}}_2^2$ is smooth and its gradient function is Lipschitz continuous with the Lipschitz constant $l= \lambda_{\text{max}}(\mathbf \Phi^T \mathbf \Phi)$, where $\lambda_{\text{max}} (\cdot)$ denotes the maximum eigenvalue of a matrix. Thus, the function $h(\mathbf x)=\frac{l}{2} \norm{\mathbf x}_2^2- \frac{1}{2}\norm{\mathbf y - \mathbf {\Phi x}}_2^2$ for $l= \lambda_{\text{max}}(\mathbf \Phi^T \mathbf \Phi)$ is convex.
	\end{theorem}	 
	\emph{Proof}: See Appendix B.

	Writing the $\ell_2$-square terms in \eqref{obj_proj_close_dc} by standard quadratic forms, we have 
	\begin{IEEEeqnarray*}{lCl}
	\label{quadratic_obj_proj_close_dc}
	\mathop {\text{min}}_\mathbf x 
	&& \quad \underbrace{\frac{l}{2} \mathbf x^T \mathbf x 
	+ \rho \norm {\mathbf x}_1}_{f(\mathbf x)} \\
	&& - \underbrace{\left( \frac{l}{2} \mathbf x^T \mathbf x - \frac{1}{2} \mathbf x^T \mathbf \Phi ^T \mathbf \Phi \mathbf x + ( \mathbf \Phi ^T \mathbf y )^T \mathbf x
	 + \rho \norm{\mathbf x}_{K,1}  \right) }_{g(\mathbf x)}.\\ 
	\IEEEyesnumber
	\end{IEEEeqnarray*}	
	We split the positive and negative parts of $\mathbf x$ by letting $\mathbf u = (\mathbf x)_+, \mathbf v = (-\mathbf x)_+$. By denoting $\mathbf{z} = [\mathbf{u}^{T}, \mathbf{v}^{T}]^{T}$, we express \eqref{quadratic_obj_proj_close_dc} as 
    \begin{IEEEeqnarray*}{lCl}
	\label{dc_bcqp_closed}	
	\mathop{\text{min}}_{\mathbf z}	 
	\quad &&  \underbrace{\frac{l}{2} \mathbf z^T \mathbf z + \rho \mathbf 1^T \mathbf z }
	_{f(\mathbf z)} 
	- \underbrace{ 
	\left( \frac{l}{2} \mathbf z^T \mathbf z 
	- \frac{1}{2} \mathbf{z}^{T} \mathbf{B} \mathbf{z} + \mathbf{q}^{T} \mathbf{z} 
	+ \rho \mathbf 1_K^T \mathbf z \right) 
	}
	_{g(\mathbf z)}\\
	 \text{s.t.} \quad
	 &&\mathbf z \succeq  \mathbf 0
	\IEEEyesnumber  
	\end{IEEEeqnarray*} 
where 
    \begin{IEEEeqnarray*}{lCl}
    \mathbf{z} = \left[\begin{matrix}
    \mathbf{u}\\
    \mathbf{v}
    \end{matrix}\right],
    \quad
    \mathbf{B} = \left[
    \begin{matrix}
    \mathbf{\Phi}^T \mathbf{\Phi} & -\mathbf{\Phi}^T \mathbf{\Phi}\\
    - \mathbf{\Phi}^T \mathbf{\Phi} & \mathbf{\Phi}^T \mathbf{\Phi}
    \end{matrix} \right],
    \quad
    \mathbf{q} = \left[\begin{matrix}
    \mathbf{\Phi}^T \mathbf{y}\\
    -\mathbf{\Phi}^T \mathbf{y}
    \end{matrix}\right]
    \end{IEEEeqnarray*}
	and $ \mathbf 1_K $ is an indicator vector having either one-valued or zero-valued elements, and the one-valued elements of $ \mathbf 1_K$ indicate the $K$ largest elements of $\mathbf z$. 

	Following the DC algorithm framework to solve the problem \eqref{dc_bcqp_closed}, we perform the following two steps repeatedly until convergence:
	\begin{IEEEeqnarray*}{lCl}
	\label{dc_proj_closed}
	(\text{a}) \quad 
	\partial g(\mathbf z^{t-1}) 
	= l \mathbf z^{t-1} 
	- \mathbf B \mathbf z^{t-1} + \mathbf q
	+ \rho \mathbf 1_K^{t-1} \\	
	(\text{b}) 
	\quad \mathbf z^{t} =
	\mathop{\text{argmin}}_{\mathbf z \ge \mathbf 0} 
	\lbrace \frac{l}{2} \mathbf z^T \mathbf z +\rho \mathbf 1^T \mathbf z
	- \mathbf z^T \partial g(\mathbf z^{t-1}) 
	 \rbrace
	\IEEEyesnumber  
	\end{IEEEeqnarray*} 
where the superscript $t-1$ and $t$, respectively, indicate the $(t-1)$th and the $t$th update. 
	The subproblem (b) of \eqref{dc_proj_closed} has a closed-form optimal solution. To derive it, we rewrite the subproblem (b) of \eqref{dc_proj_closed} as
	\begin{IEEEeqnarray*}{lCl}
	\label{prob_closed_z}
	\mathop{\text{min}}_{\mathbf z} 
	&& \quad
	\frac{l}{2} \mathbf z^T \mathbf z+\rho {\mathbf z}^T \mathbf 1
	- \mathbf z^T \partial g(\mathbf z^{t-1})
	 \\
	\text{s.t.} && \quad \mathbf z \succeq \mathbf{0}
	\IEEEyesnumber  
	\end{IEEEeqnarray*} 
	where 
	$\partial g(\mathbf z^{t-1}) = l \mathbf z^{t-1} - \mathbf B \mathbf z^{t-1} + \mathbf q + \rho \mathbf 1_K^{t-1}.$
	Next, the problem \eqref{prob_closed_z} can be expressed as
	\begin{IEEEeqnarray*}{lCl}
	\label{proj_prob_closed_z}
	\mathop{\text{min}}_{\mathbf z} 
	&& \quad
	\frac{l}{2} \norm{\mathbf z - \frac{1}{l} \left(\partial g(\mathbf z^{t-1}) - \rho \mathbf 1\right) }_2^2\\
	\text{s.t.} 
	&& \quad 
	\mathbf z \succeq \mathbf 0.
	\IEEEyesnumber  
	\end{IEEEeqnarray*} 
	The objective of problem \eqref{proj_prob_closed_z} is to minimize the $\norm{\mathbf z - \frac{1}{l}\left(\partial g(\mathbf z^{t-1}) - \rho \mathbf 1\right) }^2$ over $\mathbf z \succeq \mathbf 0$ with respect to variable $\mathbf z$. The solution is simply the Euclidean projection of $\frac{1}{l}\left(\partial g(\mathbf z^{t-1}) - \rho \mathbf 1\right)$ onto the nonnegative orthant. Thus, the gradient projection descent update has a closed-form optimal solution 
	\begin{IEEEeqnarray*}{lCl}
	\label{optimal_z}
	\mathbf z^* 
	&&= \textit{Proj} \left( \frac{1}{l}\left(\partial g(\mathbf z^{t-1}) - \rho \mathbf 1\right)\right) \\
	&&  =\left( \frac{1}{l}\left(\partial g(\mathbf z^{t-1}) - \rho \mathbf 1\right) \right)_+
	\IEEEyesnumber  
	\end{IEEEeqnarray*} 
	where the $\textit{Proj}(\cdot)$ represents the Euclidean projection onto the feasible set $\mathbf z \succeq \mathbf 0$, which is simply the positive-taking operation $(\cdot)_+$.
	Applying the closed-form solution \eqref{optimal_z} to the subproblem (b) of \eqref{dc_proj_closed}, the DC programming procedure in \eqref{dc_proj_closed} is simplified to perform the following two single-step computations repeatedly until convergence:
	\begin{IEEEeqnarray*}{lCl}
	\label{dc_proj_closed_simple}
	(\text a) \quad  
	\partial g(\mathbf z^{t-1})  =  l \mathbf z^{t-1} - \mathbf B \mathbf z^{t-1} + \mathbf q + \rho \mathbf 1_K^{t-1} \\	
	(\text {b}) 
	\quad \mathbf z^{t} 
	 = \left( \frac{1}{l} (\partial g(\mathbf z^{t-1}) - \rho \mathbf 1) \right)_+.
	\IEEEyesnumber  
	\end{IEEEeqnarray*} 
	We summarize the updates \eqref{dc_proj_closed_simple} as a single-loop DC gradient projection algorithm in Algorithm 2.
	By avoiding the inner-loop iterations, the computational complexity of Algorithm 2 for each iteration is largely reduced compared with Algorithm 1. 
	However, we comment that if the parameter $l$ is large, Algorithm 2 can be slow to converge. 
	Therefore, we will further propose an accelerated algorithm in the following section.

\begin{algorithm}
	\caption{Single-loop DC-GPSR-Basic (SlDC-GPSR-Basic)} 
	\hspace*{\algorithmicindent} \textbf{Input:} measurements $\mathbf y$, measurement matrix $\mathbf \Phi$ and a small number $\epsilon$ \\
    \hspace*{\algorithmicindent} \textbf{Output:} reconstruction $\hat{\mathbf x}$ \\
    	\hspace*{\algorithmicindent} \textbf{Initialization:}  $\mathbf u^0$, $\mathbf v^0$, $\mathbf z^0 \leftarrow [(\mathbf u^0)^T,  (\mathbf v^0)^T]^T$
	\begin{algorithmic}[1]
		\For {$t=1,2,\ldots$}
			\State Compute the gradient $\partial g(\mathbf z^{t-1}) = l \mathbf z^{t-1} - \mathbf B \mathbf z^{t-1} + \mathbf q + \rho \mathbf 1_K^{t-1} $ as (a) of \eqref{dc_proj_closed_simple}.
				\State Perform the optimal projection operation $\mathbf z^{t} = \left( \frac{1}{l} (\partial g(\mathbf z^{t-1}) - \rho \mathbf 1) \right)_+$ as (b) of \eqref{dc_proj_closed_simple}.
			\State Check terminate condition $\norm{\mathbf z^{t}-\mathbf z^{t-1}}_2 \le \epsilon$, return to Step 1 if not satisfied; otherwise, terminate with $\mathbf z^t =  [(\mathbf u^t)^T,  (\mathbf v^t)^T]^T$, and return the reconstruction $\hat{\mathbf x} = \mathbf u^t - \mathbf v^t$.
		\EndFor
	\end{algorithmic} 
\end{algorithm}

\subsection{An Extension Algorithm for Single-loop DC-GPSR Using Monotonic BB Step Size}
	A crucial observation on SlDC-GPSR-Basic (Algorithm 2) is that we can interpret it as a simple gradient projection descent method to solve a nonconvex optimization problem with a guarantee of global convergence.
	This observation can be obtained by substituting $\partial g(\mathbf z^{t-1})$ in (a) into (b) of \eqref{dc_proj_closed_simple} such that Step 2 and Step 3 in Algorithm 2 can be combined to be the $t$th update of $\mathbf z^t$ 
	\begin{IEEEeqnarray*}{lCl}
	\label{interpret_t_update}
	\mathbf z^t && = \textit{Proj} \left( \mathbf z^{t-1} - \frac{1}{l}\left( \mathbf B \mathbf z^{t-1} - \mathbf q - \rho \mathbf 1_K^{t-1}	+ \rho \mathbf 1 \right) \right) 
				\\	  && = \left( \mathbf z^{t-1} - \frac{1}{l} \left( \mathbf B \mathbf z^{t-1} - \mathbf q - \rho \mathbf 1_K^{t-1}	+ \rho \mathbf 1 \right) \right)_+.
	\IEEEyesnumber  
	\end{IEEEeqnarray*} 
    Thus, we can clearly see that the SlDC-GPSR-Basic is just a gradient projection descent method to solve the following nonconvex problem with the required step size $1/l$ for $l= \lambda_{\text{max}}(\mathbf \Phi^T \mathbf \Phi)$ 
    \begin{IEEEeqnarray*}{lCl}
	\label{bcqp_closed}	
	 \mathop{\text{min}}_{\mathbf z}	 
	&&\quad \frac{1}{2} \mathbf{z}^{T} \mathbf{B} \mathbf{z} - \mathbf{q}^{T} \mathbf{z} 
	- \rho \mathbf 1_K^T \mathbf z + \rho \mathbf 1^T \mathbf z
	:= F(\mathbf z)\\
	 \text{s.t.} 
	 && \quad \mathbf z \succeq  \mathbf 0
	\IEEEyesnumber  
	\end{IEEEeqnarray*} 
where $\mathbf z$, $\mathbf B$ and $\mathbf q$ have been defined in \eqref{dc_bcqp_closed}.
	The problem \eqref{bcqp_closed} is an equivalent form of the sparse reconstruction problem \eqref{dc_penalty_function}, but it uses a nonnegative double-sized variable $\mathbf z = [(\mathbf x)_+^T, (-\mathbf x)_+^T]^T$ to express the sparse vector $\mathbf x$.
	The gradient projection descent is a mature method for convex optimizations. However, its convergence is unguaranteed for solving a nonconvex problem. 
	Thus, in general, the gradient projection descent method cannot be directly applied to a nonconvex problem.
    Interestingly, as shown, the proposed SlDC-GPSR-Basic algorithm can be viewed as the gradient projection descent method applied to nonconvex optimizations.
     Since this algorithm is derived from DC programming, it carries the same global convergence property as the other DC programming algorithms \cite{Thi2018DC}.

    Although the fixed step size $1/l$ ensures a global convergence for the SlDC-GPSR-Basic algorithm, this step size can be too small in practice. 
    Thus, we design a variable step size to accelerate the single-loop DC-GPSR algorithm.
    We denote $\alpha^{t}$ by a variable step size for the $(t+1)$th update, and extend the update \eqref{interpret_t_update} to be a gradient projection descent update for the problem \eqref{bcqp_closed} with step size $\alpha^{t}$, i.e.,
    \begin{IEEEeqnarray*}{lCl}
    \label{gpsr_general}
    \tilde{\mathbf z}^{t+1} && = \textit{Proj} \left( \mathbf z^{t} - \alpha^{t} \left( \nabla F(\mathbf z^{t})\right)\right) 	\\
				  && = \left( \mathbf z^{t} - \alpha^{t} \left( \mathbf B \mathbf z^{t} - \mathbf q - \rho \mathbf 1_K^{t}	+ \rho \mathbf 1 \right) \right)_+
	\IEEEyesnumber
    \end{IEEEeqnarray*}
    where $\alpha^{t}$ can be explicitly calculated by the BB step size method by 
    \begin{IEEEeqnarray*}{lCl}
    \label{BB_step_size}
    \alpha^{t} = \frac{\norm{\mathbf z^{t}- \mathbf z^{t-1}}^2}{(\mathbf z^{t} - \mathbf z^{t-1})^T \left( F(\mathbf z^t) - F(\mathbf z^{t-1}) \right)}.
    \IEEEyesnumber
    \end{IEEEeqnarray*}
    To prevent the step size being overly large, we employ a scaler $\beta^{t+1} \in (0, 1]$ to limit the update so that the objective will descend monotonically
    \begin{IEEEeqnarray*}{lCl}
    \label{gpsr_general_2step}
    \mathbf{z}^{t+1} &= \mathbf{z}^{t} +\beta^{t} ( \tilde{\mathbf{z}}^{t+1}  -  \mathbf{z}^{t} ).
    \IEEEyesnumber
    \end{IEEEeqnarray*}
    We can calculate $\beta^{t}$ to minimize the objective $F(\mathbf{z}^{t+1})$ by
    \begin{IEEEeqnarray*}{lCl}
    \label{step_size_beta}
    \beta^{t} = \frac{(\bm \delta^{t})^T \nabla F(\mathbf z^t)}{(\bm \delta^{t})^T \mathbf B \bm \delta^t}
     \IEEEyesnumber
    \end{IEEEeqnarray*}
    where $\bm \delta^t = \tilde{\mathbf{z}}^{t+1}  -  \mathbf{z}^{t}$.
    We summarize the updates \eqref{gpsr_general} and \eqref{gpsr_general_2step} in Algorithm 3 and name it SlDC-GPSR-BB, which is an extension of single-loop DC-GPSR algorithm since it extends Algorithm 2 by introducing the BB step size.
  	\begin{algorithm}
	\caption{Single-loop DC-GPSR with monotonic BB step size (SlDC-GPSR-BB)} 
	\hspace*{\algorithmicindent} \textbf{Input:} measurements $\mathbf y$, measurement matrix $\mathbf \Phi$ and a small number $\epsilon$ \\
    \hspace*{\algorithmicindent} \textbf{Output:} reconstruction $\hat{\mathbf x}$ \\
    	\hspace*{\algorithmicindent} \textbf{Initialization:}  $\mathbf u^0$, $\mathbf v^0$, $\mathbf z^0 \leftarrow [(\mathbf u^0)^T,  (\mathbf v^0)^T]^T, \alpha^0$
	\begin{algorithmic}[1]
		\For {$t=0, 1, 2,\ldots$}
			\State Compute update $\tilde{\mathbf z}^{t+1}$ using \eqref{gpsr_general}.
			\State Compute step size $\beta^t$ using \eqref{step_size_beta}.
			\State Compute update $\mathbf z^{t+1}$ using \eqref{gpsr_general_2step}.
			\State Compute step size $\alpha^{t+1}$ using \eqref{BB_step_size}.
			\State Check terminate condition $\norm{\mathbf z^{t+1}-\mathbf z^{t}}_2 \le \epsilon$, return to Step 1 if not satisfied; otherwise, terminate with $\mathbf z^{t+1} =  [(\mathbf u^{t+1})^T,  (\mathbf v^{t+1})^T]^T$ and return the reconstruction $\hat{\mathbf x} = \mathbf u^{t+1} - \mathbf v^{t+1}$.
		\EndFor
	\end{algorithmic} 
\end{algorithm}

\bb{
\section{Complexity Analysis}
	Table \ref{complexity} shows the computational cost of the main computing operations and compares the computational complexities of the proposed algorithms. 
	Common operations for all three algorithms mainly contain matrix-vector multiplications, vector inner products, vector sums, scaler-vector multiplications and the subgradient of the top-$(K,1)$ norm for a nonnegative vector.
	Among these operations, the computationally intensive terms have the subgradient computation $\partial \norm{\mathbf z}_{K, 1}$ and the matrix-vector product $\mathbf {Bz}$ whose computational costs are shown in Table \ref{complexity}.
	To analyze the overall time complexities of the proposed algorithms, we consider both the computational costs at each iteration and the convergence rates. 
	The computational costs at each iteration \footnote{To compare fairly, one iteration is considered from Step 2 to Step 5 for DlDC-GPSR, Step 2 and Step 3 for SlDC-GPSR-Basic and from Step 2 to Step 5 for SlDC-GPSR-BB.} for these algorithms are shown in Table \ref{complexity}, while the convergence rates will be illustrated in Section VII.B.
	By taking into account of the convergence rates, we can conclude that, among the three proposed algorithms, SIDC-GPSR-BB has the lowest time complexity. 
	
	\begin{table*}[tbp]
	\scriptsize
	\center
	\caption{\bb{Computational costs for operations (left) and computational complexity per iteration for different algorithms (right), $n$ is the dimension of the sparse vector $\mathbf x$, and $2n$ is the length of vector $\mathbf z$, $m$ is the dimension of the measurements $\mathbf y$}}
  	\label{complexity}
    \begin{tabular}{|>{\centering\arraybackslash} m{2.3cm}
    					|>{\centering\arraybackslash} m{2.3cm}
    					|>{\centering\arraybackslash} m{2.3cm}
    					||>{\centering\arraybackslash} m{2.3cm}
    					|>{\centering\arraybackslash} m{2.3cm}
    					|>{\centering\arraybackslash} m{2.3cm}
    					|} 
    \hline
 	Operation/Algorithm & $\partial \norm{\mathbf z}_{K, 1}$  & $\mathbf {Bz}$ & DlDC-GPSR  & SlDC-GPSR-Basic & SlDC-GPSR-BB \\
    \hline
  	Time complexity & $O(2n \log (2n))$ \tablefootnote{A feasible subgradient computation of $\partial \norm{\mathbf z}_{K, 1}$ requires the indices of the $K$ largest elements to be obtained by sorting the elements, which can be accomplished with the time complexity $O(2n \log (2n))$. This time complexity is only an estimate, as more efficient sorting algorithms can exist.} & $O(mn)$ & $I \times O(mn)$ \tablefootnote{ We assume that an arbitrary outer iteration of DlDC-GPSR has $I$ inner loops. Given a termination condition, the number of inner-loop iterations $I$ can vary for each outer-loop iteration of DlDC-GPSR.} & $O(mn)$ & $O(mn)$ \\
    \hline
    \end{tabular}
	\end{table*}

}
\section{Numerical Results}
\subsection{Experiment Setup and Performance Metrics}
	We consider a downlink massive MIMO system having half-wavelength spaced ULA at the BS.
	The number of BS antennas is set as $N_t=256$.
	Without loss of generality, the number of UE antennas is set as $N_r=1$ \footnote{According to the proposed column-wise broadcasting method for vectorizing beamspace channel matrix $\mathbf H \in \mathbb{C}^{N_t \times N_r}$, to reconstruct $\mathbf H$ we simply reconstruct its $N_r$ columns independently.}.
	We consider the narrowband block-fading channels, and set a channel coherence block as $L_c = 600$ symbols.
	We randomly generate $1,000$ channel vectors according to the Saleh-Valenzuela channel model described in (1), where the number of paths is set as $N_p = 3$. 
	For each path, the complex channel gain $\alpha_l$ follows a complex Gaussian distribution; the AoA and AoD, i.e., $\theta_{r, l}$ and $\theta_{t, l}$, are uniformly distributed over $[-\pi/2, \pi/2]$.
	We transform the generated channel vectors into the angular domain using a DFT. 
	To consider power leakage, we set the number of nonzero channel coefficients as $N_{s} = 16$ by neglecting the small-value beamspace channel coefficients \footnote{Although in practice the number of nonzero channel coefficients can be unknown and uncertain, in the simulation we fix the number of nonzero channel coefficients for fair performance comparisons. We set $N_{s} = 16$ because our observations revealed that the $16$ largest-magnitude channel coefficients can cover most of the energy of a channel vector, i.e., $\norm{\mathbf h_i}_2^2$ for $1 \le i \le N_r$.}. 
	Unless stated otherwise, we adopt the random Gaussian matrix as the measurement matrix $\mathbf S_G \in \mathbb{R}^{L \times N_t}$, and the pilot matrix can be obtained by $\mathbf P = \mathbf{S}_G^T \mathbf{U}_t \in \mathbb{C}^{N_t \times L}$.
	All the simulations were implemented on a desktop computer equipped with a 3.2 GHz Intel Core i7-8700 CPU with 8GB of physical memory.

	We adopt the normalized mean squared error (NMSE) to evaluate the channel reconstruction accuracy.	
	\bb{The NMSE is defined as $\text{NMSE} = \frac{1}{N} \sum\limits_{i=1}^N \frac{\norm{\mathbf H_i - \hat{\mathbf H_i}}_F^2}{\norm{\mathbf H_i}_F^2}$, where $N$ is the number of channel samples.}
	We use achievable spectral efficiency to evaluate the influences of pilot overhead and SNR penalty caused by imperfect channel estimation on data communication. The achievable spectral efficiency $R/B$ is defined as $ R/B = (1-\alpha)C(\text{SNR}_\text{eff})$ \cite[eq. (5.195)]{Heath2018}, 
	where $R$ is the achievable rate; $B$ is the bandwidth; $\alpha$ is the ratio of pilot training in a channel coherence block duration, and it can be calculated as $L/L_c$, where $L_c$ is the length of channel coherence block; $C(\text{SNR}_\text{eff})$ is the channel capacity at $\text{SNR}_\text{eff}$, and $\text{SNR}_\text{eff}$ refers to the efficient SNR by considering imperfect channel estimation \cite{Heath2018}.
	According to the linear equation $\mathbf R = \mathbf {SH} + \mathbf W$  in \eqref{CS_multi_antenna}, we define the system SNR as $\text{SNR} = \frac{\norm{\mathbf {SH}}_F^2}{\norm{\mathbf {W}}_F^2}$,
	where $\mathbf S$ is the measurement matrix, $\mathbf H$ is the beamspace channel, and $\mathbf {W}$ is the noise. 

\subsection{Illustrations of Beamspace Channel Reconstructions and Algorithm Convergence Property}

\begin{figure*}[!tph]
\centering
\captionsetup{justification=centering}
{\includegraphics[scale=0.36]{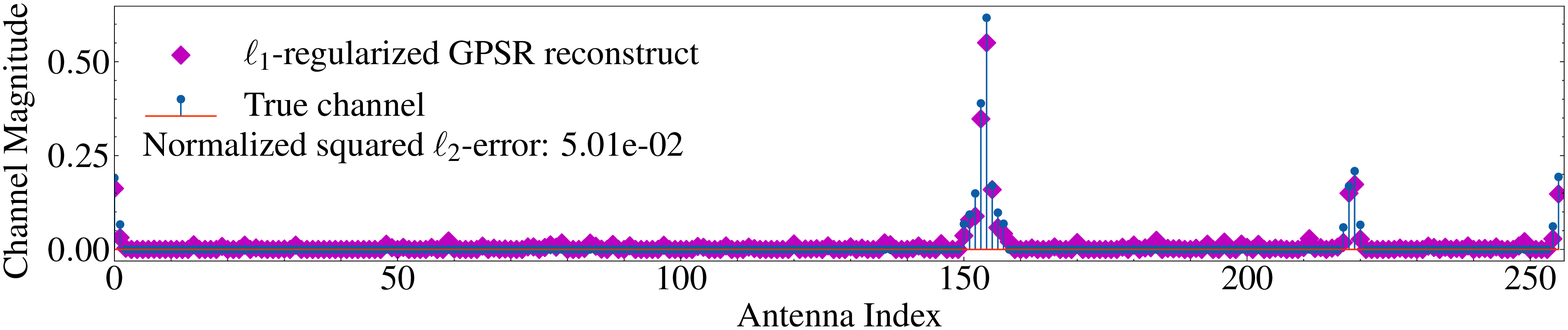}}\\
{\includegraphics[scale=0.36]{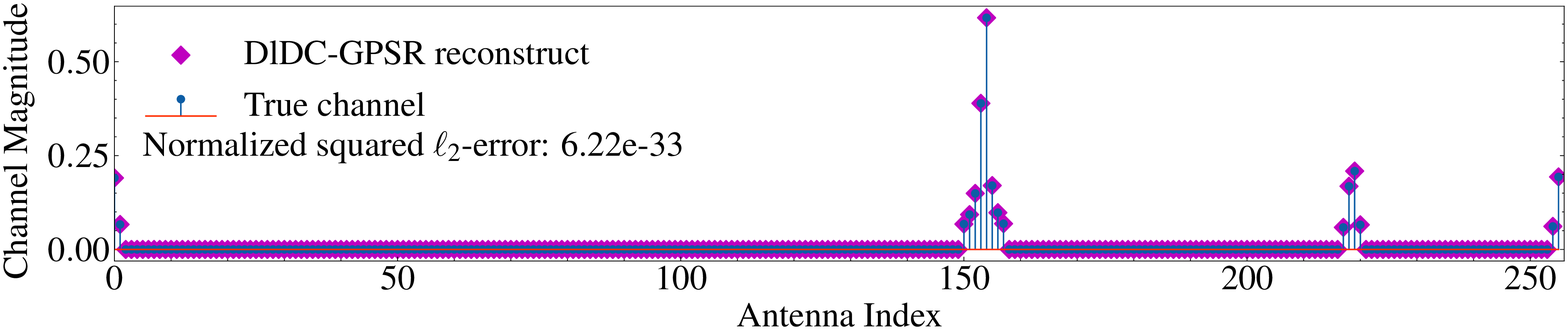}}\\
{\includegraphics[scale=0.36]{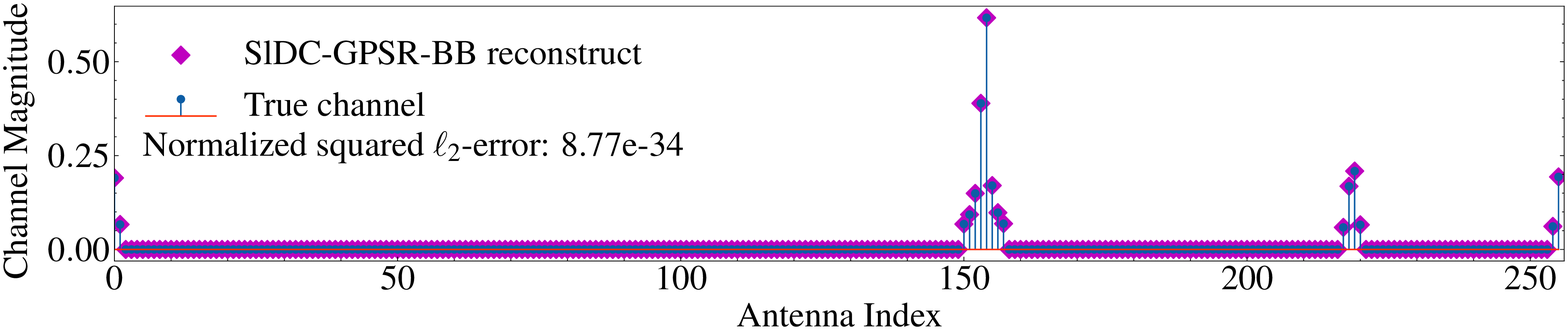}} \\
\caption{Channel magnitudes of a true channel sample $\mathbf x_\text{opt}$ and the reconstruction $\hat{\mathbf x}$ by the conventional $\ell_1$-regularized GPSR algorithm, the proposed algorithms DlDC-GPSR and SlDC-GPSR-BB. The normalized squared $\ell_2$-error is defined as $\norm{\mathbf{x}_\text{opt}-\hat{\mathbf{x}}}_2^2/\norm{\mathbf x_\text{opt}}_2^2$}
\label{fig5}
\end{figure*}	

\begin{figure}[t]
\centering
\normalsize
\includegraphics[width=3.5in]{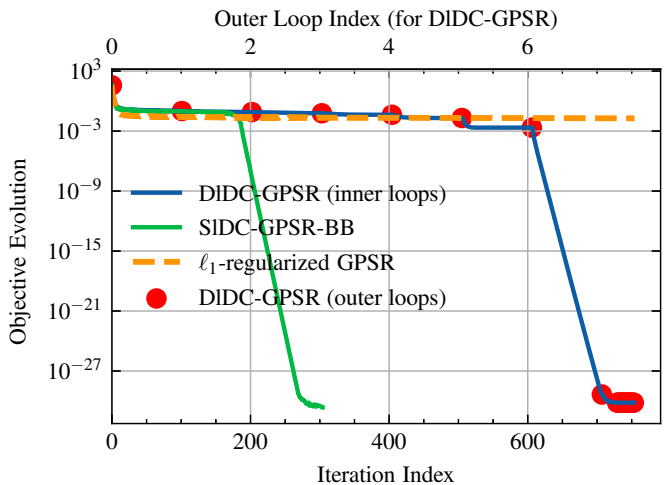}
\captionsetup{justification=centering}
\caption{Objective values versus iterations for reconstructing a sample of beamspace channel vector. The evaluating objectives are $\frac{1}{2}\norm{\mathbf{y}-\mathbf{\Phi}\mathbf{x}}_2^2 + \rho (\norm{\mathbf x}_1 - \norm{\mathbf x}_{K,1})$  for DlDC-GPSR and SlDC-GPSR-BB, and is $\frac{1}{2}\norm{\mathbf{y}-\mathbf{\Phi}\mathbf{x}}_2^2 + \rho \norm{\mathbf x}_1$ for $\ell_1$-regularized GPSR}
\label{fig1_4_fig1}
\end{figure}

\begin{figure}[t]
\centering
\normalsize
\includegraphics[width=3.5in]{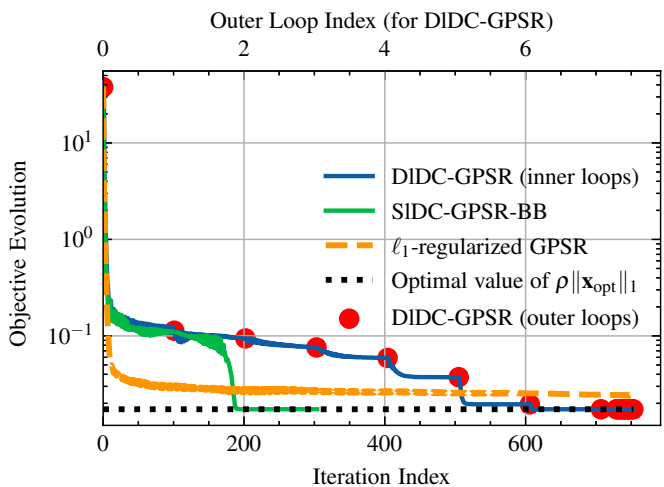}
\captionsetup{justification=centering}
\caption{Objective values ($\frac{1}{2}\norm{\mathbf{y}-\mathbf{\Phi}\mathbf{x}}_2^2 + \rho \norm{\mathbf x}_1$) versus iterations for the reconstruction of a sample of beamspace channel vector}
\label{fig1_4_fig2}
\end{figure}

\begin{figure}[t]
\centering
\normalsize
\includegraphics[width=3.5in]{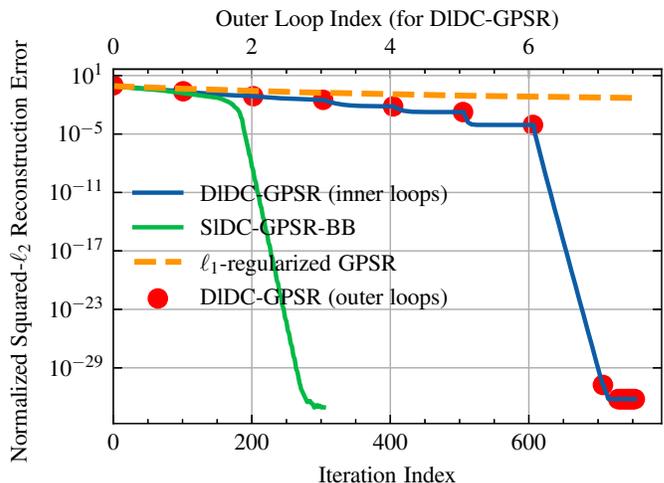}
\captionsetup{justification=centering}
\caption{Reconstruction error ($\norm{\hat{\mathbf x}-\mathbf x_\text{opt}}_2^2 / \norm{\mathbf x_\text{opt}}_2^2 )$ versus iterations for the reconstruction of a sample of beamspace channel vector}
\label{fig1_4_fig3}
\end{figure}

	In this subsection, we evaluate the reconstruction accuracies and convergence rates of DlDC-GPSR (Algorithm 1) and SlDC-GPSR-BB (Algorithm 3) \footnote{As we have shown, the SlDC-GPSR-Basic (Algorithm 2) has an interesting theoretical interpretation as a pure gradient projection algorithm, but the required step size $1/l$ is often too small to converge within a reasonable period. Therefore, we do not consider Algorithm 2 in simulations.} by reconstructing an arbitrary channel vector sample $\mathbf x$ and compare the performances with conventional $\ell_1$-regularized GPSR algorithm \cite{figueiredo2007gradient}.
	For fair comparisons, we adopt BB step sizes for the inner loops of DlDC-GPSR.
	The pilot length is set as $L=128$.	
	\mbox{Fig. \ref{fig5}} shows the magnitudes of the true sparse beamspace channel and the reconstructions.
	We can see that the DlDC-GPSR and SlDC-GPSR-BB algorithm can achieve perfect reconstructions with normalized squared $\ell_2$-errors $6.22\times10^{-33}$ and $8.77\times10^{-34}$, whereas the \mbox{$\ell_1$-regularized} GPSR reconstruction has noticeable errors with the error $5.01 \times 10^{-2}$.
	By setting the termination condition as $\norm{\hat{\mathbf x}^t -  \hat{\mathbf x}^{t-1}}_2 \le 10^{-30}$, the runtimes are about $0.08$, $0.08$ and $0.05$ seconds for \mbox{$\ell_1$-regularized} GPSR, DlDC-GPSR and SlDC-GPSR-BB, respectively.
	To show their convergence properties within runtimes, we plot objective values versus iterations in \mbox{Fig. \ref{fig1_4_fig1}}.
	It should be noted that the objective of the proposed DlDC-GPSR and SlDC-GPSR-BB algorithms is $\frac{1}{2}\norm{\mathbf{y}-\mathbf{\Phi}\mathbf{x}}_2^2 + \rho (\norm{\mathbf x}_1 - \norm{\mathbf x}_{K,1})$, whereas the objective of conventional $\ell_1$-regularized GPSR is $\frac{1}{2}\norm{\mathbf{y}-\mathbf{\Phi}\mathbf{x}}_2^2 + \rho \norm{\mathbf x}_1$.
	 For DlDC-GPSR algorithm, the red dots indicate objective values for outer iterations, and the blue solid line shows the objective values along all inner iterations.
	We can see the proposed DlDC-GPSR algorithm decreases its objective significantly after the sixth \mbox{outer-step} and approaches the optimal value at the eighth step.
	The SlDC-GPSR-BB algorithm has the fastest convergence and can achieve almost the same objective value as the DlDC-GPSR algorithm.
	On the contrary, the objective of $\ell_1$-regularized GPSR converges to a relatively large value.
	\mbox{Fig. \ref{fig1_4_fig2}} shows the evolution of $\ell_1$-norm penalized least-square objective values, i.e., $\norm{\mathbf y - \mathbf \Phi \mathbf x}_2^2 + \rho \norm{\mathbf x}_1$.
	In \mbox{Fig. \ref{fig1_4_fig2}}, we also plot the optimal values of the penalty term $\rho \norm{\mathbf x_{\text{opt}}}_1$, where $\mathbf x_{\text{opt}}$ represents the true sample of a beamspace channel vector.
	We can see that the DlDC-GPSR and SlDC-GPSR-BB algorithms can arrive and stay at the objective value $\rho \norm{\mathbf x_{\text{opt}}}_1$, which is the optimal value that the objective $\norm{\mathbf y - \mathbf \Phi \mathbf x}_2^2 + \rho \norm{\mathbf x}_1$ can reach when $\norm{\mathbf y - \mathbf \Phi \hat{\mathbf x}}_2^2 = 0$ and $\rho \norm{\hat{\mathbf x}}_1 = \rho \norm{\mathbf x_{\text{opt}}}_1$. 
	However, the \mbox{$\ell_1$-regularized} GPSR cannot achieve this optimal value and has a noticeable gap from the optimal value.
	It is meaningful to observe this minimum objective gap between our proposed algorithms and the conventional $\ell_1$-regularized GPSR algorithm, because this gap can provide important insight into the approximation error introduced by relaxing the $\ell_0$-norm as $\ell_1$-norm.
	\mbox{Fig. \ref{fig1_4_fig3}} shows the normalized squared-$\ell_2$ errors of reconstruction versus the number of iterations.
	We observe that both the DlDC-GPSR and SlDC-GPSR-BB algorithm can achieve accurate reconstructions having errors on the order of $10^{-33}$ and $10^{-34}$, which is far more accurate than the reconstruction by the conventional $\ell_1$-regularized GPSR algorithm having an error on the order of $10^{-2}$.

\vspace{-4mm}
\subsection{Performance Comparisons With Other Algorithms}	
	We compare the reconstruction performances of the proposed algorithms with several existing sparse reconstruction algorithms including the $\ell_1$-regularized GPSR \cite{figueiredo2007gradient}, ISTA \cite{blumensath2008thomas}, and OMP \cite{tropp2007signal}, as well as the existing state-the-art channel estimation schemes including the SD \footnote{The SD algorithm is the support detection-based channel estimation algorithm \cite{Dai2016Beamspace}.} \cite{Dai2016Beamspace} and the SCAPMI algorithms \cite{Yang2018TCOM}.

\begin{figure}[htp]
\centering
\includegraphics[width=3.5in]{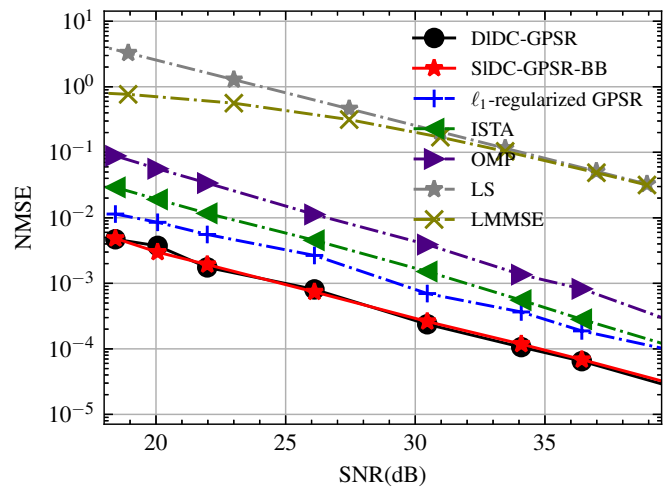}
\caption{Channel reconstruction NMSE versus SNR (dB) with training pilot length $L=128$}
\label{plot_nmse_snr1}
\end{figure}
	
	In Fig. \ref{plot_nmse_snr1}, we compare the channel reconstruction NMSE with conventional channel estimation methods including the LS and LMMSE, and popular sparse reconstruction algorithms including the OMP, ISTA and $\ell_1$-regularized GPSR.
	For conventional LS and LMMSE channel estimation methods, we use the optimal pilot matrix that contains orthonormal training pilot sequences having the length of $L = 256$.
	For the sparse reconstruction algorithms including DlDC-GPSR, SlDC-GPSR-BB, OMP, ISTA and $\ell_1$-regularized GPSR, we set the training pilot length as $L = 128$.
	Fig. \ref{plot_nmse_snr1} shows that the proposed DlDC-GPSR and SlDC-GPSR-BB algorithms achieve approximately the same accuracy. 
	Even consuming more training pilots, both LS and LMMSE channel estimation methods have higher channel reconstruction errors than the sparse reconstruction methods.
	Also, the proposed algorithms DlDC-GPSR and \mbox{SlDC-GPSR-BB} have higher accuracies than the other sparse reconstruction algorithms.
	The average runtimes required for reconstructing a channel sample are summarized in Table \ref{time} for different sparse reconstruction algorithms. 
	
	\begin{table*}[tbp]
	\scriptsize
	\centering
  	\caption{Comparison of the average runtimes per channel sample reconstruction}
  	\label{time}
    \begin{tabular}{|>{\centering\arraybackslash} m{2.3cm}
    					|>{\centering\arraybackslash} m{2.3cm}
    					|>{\centering\arraybackslash} m{2.3cm}
    					|>{\centering\arraybackslash} m{2.5cm}
    					|>{\centering\arraybackslash} m{2.3cm}
    					|>{\centering\arraybackslash} m{2.3cm}
    					|} 
    \hline
 	Algorithm & SlDC-GPSR-BB  & DlDC-GPSR & $\ell_1$-regularized GPSR & ISTA & OMP \\
    \hline
  	$\text{SNR}=40$ dB & 0.03s  & 0.67s & 0.12s & 0.61s & 0.16s \\
    \hline
  	$\text{SNR}=30$ dB & 0.07s & 1.61s & 0.19s &  0.86s &  0.21s \\
    \hline
 	$\text{SNR}=18$ dB & 0.16s & 2.39s & 0.26s & 1.89s & 0.31s\\
    \hline
    \end{tabular}
	\end{table*}

\begin{figure}[htp]    
\centering
\includegraphics[width=3.3in]{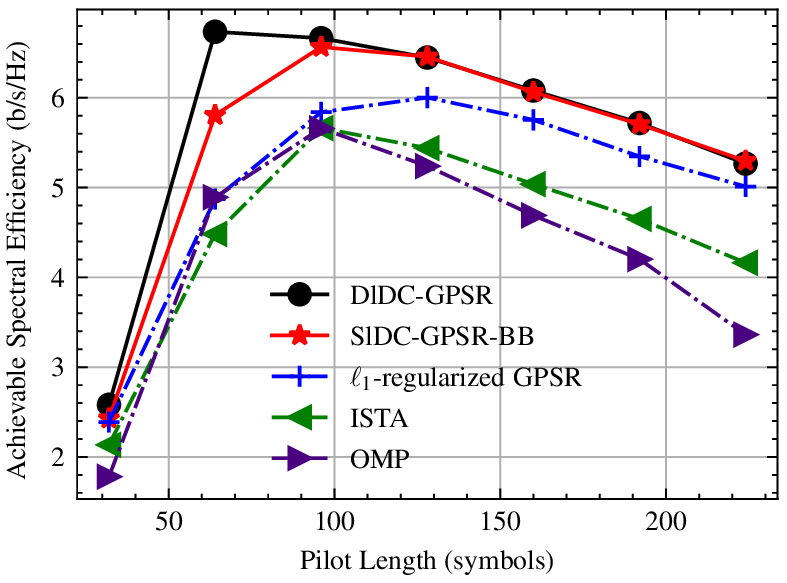}
\caption{Achievable spectral efficiency $R/B$ versus training pilot length $L$ at $\text{SNR}=25$ dB}
\label{plot_RateLen_SNR25}
\end{figure}

\begin{figure}[htp]    
\centering
\includegraphics[width=3.3in]{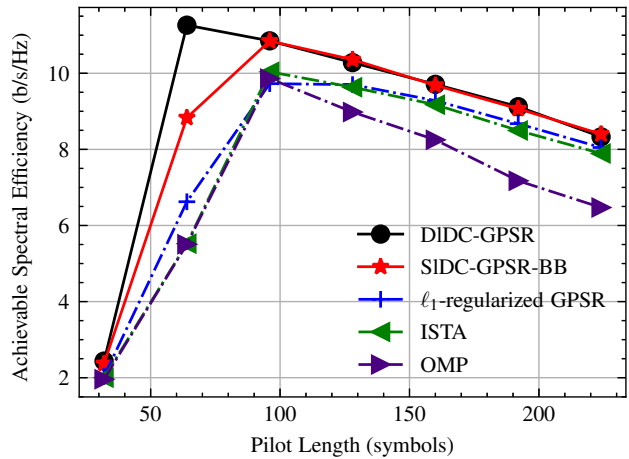}
\caption{Achievable spectral efficiency $R/B$ versus training pilot length $L$ at $\text{SNR}=40$ dB}
\label{plot_RateLen_SNR40}
\end{figure}

	For sparse channel reconstructions, increasing the pilot length can improve reconstruction accuracy. 
	However, longer pilot length will also increase the training overhead and subsequently degrade the spectral efficiency. 
	In other words, there exists a tradeoff between the training pilot overhead and spectral efficiency.
	To illustrate this tradeoff and comparison for different sparse reconstruction algorithms, we plot the achievable spectral efficiency versus training pilot length $L$ for $\text{SNR}=25$ dB and $\text{SNR}=40$ dB in Fig. \ref{plot_RateLen_SNR25} and Fig. \ref{plot_RateLen_SNR40}, where the length of a channel coherence block is set as $L_c = 600$ symbols, and the training pilot length is set as $L \in \{32, 64, ..., 256\}$.
	We can see that the achievable spectral efficiencies increase when the training pilot length $L$ increases, and then decrease after reaching the peak values.
	Since when $L$ has a small value, the channel estimation error can be decreased by increasing the number of training symbols, such that the efficient SNR ($\text{SNR}_\text{eff}$) increases and dominates the achievable spectral efficiency.
	However, when $L$ is sufficiently large, a further increase of $L$ will give a diminishing benefit while the data communication ratio $(1-\alpha)$ decreases linearly, where $\alpha = L/L_c$ is the ratio of pilot training overhead in a coherence block of fading channels.
	Fig. \ref{plot_RateLen_SNR25} and Fig. \ref{plot_RateLen_SNR40} show that the proposed DlDC-GPSR and SlDC-GPSR-BB algorithms can attain higher achievable spectral efficiencies than the $\ell_1$-regularized GPSR, ISTA and OMP sparse reconstruction algorithms. 
\begin{figure}[thp]
\centering
\includegraphics[width=3.3in]{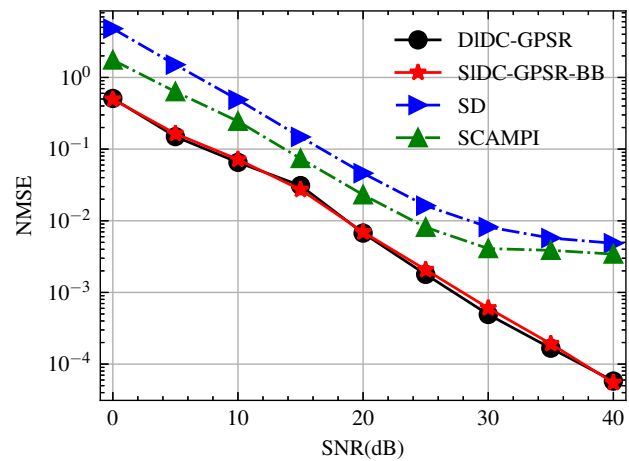}
\caption{NMSE comparison between the SCAMPI algorithm under uniform distribution with a lens antenna array in the size of $16 \times 16$, the SD algorithm and the proposed SlDC-GPSR-BB and DlDC-GPSR algorithms with a ULA in the size of $256$ while using a training pilot length of $L=96$}
\label{plot_nmse_snr2}
\end{figure}
\begin{figure}[thp]
\centering
\includegraphics[width=3.2in]{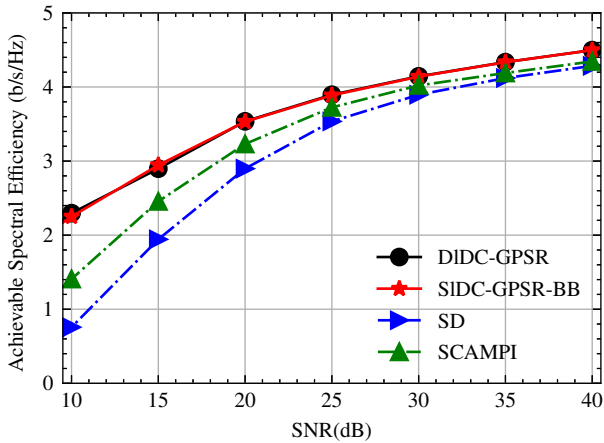}
\caption{Comparison of the achievable spectral efficiency $R/B$ between the SCAMPI algorithm under uniform distribution with a lens antenna array in the size of $16 \times 16$, the SD algorithm and the proposed SlDC-GPSR-BB and DlDC-GPSR algorithms with a ULA in the size of $256$ while using a training pilot length of $L=96$}
\label{plot_rate_snr}
\end{figure}
	We now compare the performances of the proposed algorithms with the state-of-the-art beamspace channel estimation methods including the SD and the SCAMPI algorithms.
	For fair comparisons, we selected a $16 \times 16$ lens antenna array size for SCAMPI, and chose $N_t = 256$ for the ULA antenna number for both the SD and the proposed algorithms.
	The training pilot length is set as $L=96$.
	To allow the proposed algorithms to be contrasted fairly with the SD and SCAMPI algorithms, we constructed the same type of measurement matrix used in \cite{Dai2016Beamspace} and \cite{Yang2018TCOM}. 
	The measurement matrix is a Rademacher matrix, which contains entries drawn from $\{+1, -1\}$ with equal probabilities. 
	\mbox{Fig. \ref{plot_nmse_snr2}} shows the results of channel reconstruction NMSE versus SNR. 
	At high SNR, the SD and SCAMPI algorithms appear to reach an error floor of approximately $10^{-2}$, while the proposed SlDC-GPSR-BB and DlDC-GPSR algorithms achieve lower reconstruction errors.
	\mbox{Fig. \ref{plot_rate_snr}} compares the achievable spectral efficiencies by different channel estimation schemes.
	The proposed SlDC-GPSR-BB and DlDC-GPSR algorithms achieve higher spectral efficiency than the SD and SCAMPI algorithms.

\section{Conclusion}
	We proposed three DC programming gradient projection sparse reconstruction algorithms for massive MIMO beamspace channel estimation, and they are DlDC-GPSR, SlDC-GPSR-Basic and SlDC-GPSR-BB.
	We derived these algorithms by solving a least squares problem having a nonconvex regularizer.
	The regularizer is simply the difference between an $\ell_1$-norm and a \mbox{top-$(K,1)$ norm}, which was introduced to remove the penalties on $K$ largest-magnitude elements of the reconstructing vector. 
	We employed DC programming and a gradient projection method to solve the nonconvex sparse reconstruction, and derived different double-loop and single-loop algorithms by different DC decompositions.
	The double-loop DlDC-GPSR algorithm is simple, accurate, and robust.
	The SlDC-GPSR-Basic algorithm has simple single-loop updates, and it shares the same theoretical convergence property as the other DC programming methods.
	We observed that the SlDC-GPSR-Basic updates can be perfectly interpreted as a simple gradient projection descent update.
	Based on this observation, we adopted the BB step-size strategy and proposed an extension algorithm SlDC-GPSR-BB, which has a significantly improved convergence rate and can achieve approximately the same level accuracy as the double-loop algorithm DlDC-GPSR.
	Several numerical demonstrations were provided to show the advantages of the proposed algorithms, such as high accuracies and fast computations.

\section{Acknowledgement}
	The authors thank Dr. Hui Ma for his early contribution to Algorithm 1 \cite{Wu2020sparse}.

\appendix

\subsection{Proof of Theorem 1:} 	
%
%
%

	Suppose that $\mathbf x_{\rho^*}$ is an optimal solution to \eqref{dc_penalty_function} with given $\rho^*$, then $\mathbf x_{\rho^*}$ is also optimal to \eqref{sparse_recovery3} as long as $\norm{\mathbf x_{\rho^*}}_0 \le K$ (or $\norm{\mathbf x_{\rho^*}}_1 - \norm{\mathbf x_{\rho^*}}_{K,1} = 0$) satisfies.
	Thus, we only need to prove that $\norm{\mathbf x_{\rho^*}}_0 > K$ is infeasible.
	We assume that $\mathbf x_{\rho^*}$ is an optimal solution to \eqref{dc_penalty_function} with $\rho^* > \mathop{\text{max}}_{i}\{|(\mathbf \Phi^T \mathbf y)_{i}| + q(\norm{\mathbf \Phi ^T \mathbf \Phi \mathbf e_i}_2 + |(\mathbf{\Phi}^T\mathbf{\Phi})_{ii}|/2 ) \}$.
	For $\norm{\mathbf x_{\rho^*}}_0 > K$, we construct a feasible solution to \eqref{sparse_recovery3} as $\mathbf x^\prime = \mathbf x_{\rho^*} - x_i \mathbf e_i$, where $i$ represents the index of the $(K+1)$th largest-magnitude element of vector $\mathbf x_{\rho^*}$; $\mathbf e_i$ represents the unit vector in which the $i$th element is one while the other elements are zeros; $x_i$ represents the $i$th element of vector $\mathbf x_{\rho^*}$.	
	By writing the objective of \eqref{dc_penalty_function} as 
	\begin{IEEEeqnarray*}{lCl}
	F(\mathbf x) = \frac{1}{2} \mathbf x^T \mathbf \Phi ^T \mathbf \Phi \mathbf x 
	-  (\mathbf \Phi ^T \mathbf y )^T \mathbf x
	+ \rho^* \norm{\mathbf x}_1  
	- \rho^* \norm{\mathbf x}_{K,1} \\
		\IEEEyesnumber
	\end{IEEEeqnarray*}
	we have 
	\begin{IEEEeqnarray*}{lCl}
		&& \quad  F(\mathbf x_{\rho^*}) - F(\mathbf x^\prime) \\
	  && = F(\mathbf x_{\rho^*}) - F(\mathbf x_{\rho^*} - x_i \mathbf e_i)
	 \\&& = \frac{1}{2} \mathbf x_{\rho^*}^{T} \mathbf \Phi ^T \mathbf \Phi \mathbf x_{\rho^*}-  (\mathbf \Phi ^T \mathbf y )^T \mathbf x_{\rho^*}
	 \\ && \quad - \frac{1}{2} (\mathbf x_{\rho^*} - x_i \mathbf e_i)^T \mathbf \Phi ^T \mathbf \Phi (\mathbf x_{\rho^*} - x_i \mathbf e_i) 
	 \\&& \quad +  (\mathbf \Phi ^T \mathbf y )^T (\mathbf x_{\rho^*} - x_i \mathbf e_i)
	 + \rho^* |x_i|
	 \\&& = x_i \mathbf e_i^T \mathbf \Phi^T \mathbf \Phi \mathbf x_{\rho^*}
	 -\frac{1}{2} x_i^2 (\mathbf \Phi^T \mathbf \Phi)_{i,i}
	 - x_i (\mathbf \Phi ^T \mathbf y )_i + \rho^* |x_i|
	 \\&& \ge - x_i \mathbf e_i^T \mathbf \Phi^T \mathbf \Phi \mathbf x_{\rho^*}
	 -\frac{1}{2} x_i^2 (\mathbf \Phi^T \mathbf \Phi)_{i,i}
	 - x_i (\mathbf \Phi ^T \mathbf y )_i
	 + \rho^* |x_i|
	 \\&& \ge - |x_i| \norm{\mathbf e_i^T \mathbf \Phi^T \mathbf \Phi}_2 \norm{\mathbf x_{\rho^*}}_2 
	 - \frac{1}{2} |x_i| \norm{\mathbf x_{\rho^*}}_2 |(\mathbf \Phi^T \mathbf \Phi)_{i,i}| 
	 \\ && \quad - |x_i|  \cdot |(\mathbf \Phi ^T \mathbf y )_i|
	 + \rho^* |x_i|
	 \\&&= |x_i|\big(  \rho - \norm{\mathbf e_i^T \mathbf \Phi^T \mathbf \Phi}_2 \norm{\mathbf x_{\rho^*}}_2  - \frac{1}{2} \norm{\mathbf x_{\rho^*}}_2 |(\mathbf \Phi^T \mathbf \Phi)_{i,i}| 
	 \\&& \quad -|(\mathbf \Phi ^T \mathbf y )_i| \big)
	 \\&& > 0
	\IEEEyesnumber
	\end{IEEEeqnarray*}
	which contradicts the assumption that $\mathbf x_{\rho^*}$ is an optimal solution to \eqref{dc_penalty_function}.

\subsection{Proof of Theorem 2:} 	
	We first derive the Lipschitz constant $l= \lambda_{\text{max}}(\mathbf \Phi^T \mathbf \Phi)$, then we prove the convexity of $h(\mathbf x)$. 
	We denote the least squares objective by the function $l(\mathbf x)$.
	Since the least squares objective $w(\mathbf x)=\frac{1}{2}\norm{\mathbf y - \mathbf {\Phi x}}_2^2$ is smooth if and only if its gradient function is Lipschitz continuous, we assume there exists $l < \infty$, which is named as a Lipschitz constant, such that 
	\begin{IEEEeqnarray*}{lCl}
	\norm{\nabla w(\mathbf x)- \nabla w(\mathbf z)}_2 \le l \norm{\mathbf x- \mathbf z}_2.
	\IEEEyesnumber
	\end{IEEEeqnarray*}
	For $w(\mathbf x) = \frac{1}{2}\norm{\mathbf y - \mathbf {\Phi x}}_2^2$, we have
	\begin{IEEEeqnarray*}{lCl}
	\norm{\nabla w(\mathbf x)- \nabla w(\mathbf z)}_2  
	&& = \norm{\mathbf \Phi ^T (\mathbf \Phi \mathbf x - \mathbf y) - \mathbf \Phi ^T (\mathbf \Phi \mathbf z - \mathbf y)}_2 \\
	&& = \norm{\mathbf \Phi ^T \mathbf \Phi (\mathbf x - \mathbf z)}_2 \\
	&& \le \normmm{\mathbf \Phi ^T \mathbf \Phi}_2 \norm{\mathbf x - \mathbf z}_2 \\
	&& = \lambda_{max}{(\mathbf \Phi ^T \mathbf \Phi)} \norm{\mathbf x - \mathbf z}_2
	\IEEEyesnumber
	\end{IEEEeqnarray*}
	 where $\normmm{\cdot}_2$ represents the spectral norm of a matrix, and $\lambda_{max}(\cdot)$ represents the largest eigenvalue of a matrix.
	 Thus, we obtain the Lipschitz constant $l= \lambda_{\text{max}}(\mathbf \Phi^T \mathbf \Phi)$.
	
	For $h(\mathbf x) = \frac{l}{2} \norm{\mathbf x}_2^2- \frac{1}{2}\norm{\mathbf y - \mathbf {\Phi x}}_2^2$, we have the Hessian matrix $\nabla^2 h(\mathbf x)$ as
	\begin{IEEEeqnarray*}{lCl}
	 \nabla^2 h(\mathbf x) &&= \frac{\partial (\nabla h(\mathbf x)) }{\partial \mathbf x}
	 \\&& = \frac{\partial (l \mathbf x^T - \mathbf \Phi ^T (\mathbf \Phi \mathbf x- \mathbf y)) }{\partial \mathbf x}
	 \\&& = l \mathbf I - \mathbf \Phi^T \mathbf \Phi
	\IEEEyesnumber
	\end{IEEEeqnarray*}
	where $\mathbf I$ is the identity matrix.
	 For $l= \lambda_{\text{max}}(\mathbf \Phi^T \mathbf \Phi)$, we have $l \mathbf I - \mathbf \Phi^T \mathbf \Phi \succeq 0$, which means that the Hessian matrix $l \mathbf I - \mathbf \Phi^T \mathbf \Phi $ of $h(\mathbf x)$ is semidefinite positive.
	 Thus, $h(\mathbf x)$ is a convex function of vector $\mathbf x$.

	

\bibliographystyle{IEEEtran}

\bibliography{IEEEabrv,mybib_cand}

\bibliographystyle{IEEEtran}

\end{document}